\documentclass[12pt]{amsart}

\usepackage{amsmath,amssymb,lscape}
\usepackage[all]{xy}
\usepackage{enumerate}

\newtheorem{theorem}{Theorem}[section]
\newtheorem{proposition}[theorem]{Proposition}
\newtheorem{lemma}[theorem]{Lemma}

\theoremstyle{definition}
\newtheorem{definition}[theorem]{Definition}
\newtheorem{example}[theorem]{Example}

\theoremstyle{remark}
\newtheorem{remark}[theorem]{Remark}

\numberwithin{equation}{section}

\def\DJ{{\hbox{D\kern-.8em\raise.15ex\hbox{--}\kern.35em}}}
\def\DJo{$\;$\kern-.4em
    \hbox{D\kern-.8em\raise.15ex\hbox{--}\kern.35em koori\'c}}

\font\germ=eufm10

\def\gso{{\mbox{\germ so}}}
\def\gg{{\mbox{\germ g}}}
\def\mfk{{\mbox{$\mathfrak k$}}}
\def\mfp{{\mbox{$\mathfrak p$}}}
\def\mfa{{\mbox{$\mathfrak a$}}}

\def\al{{\alpha}}
\def\be{{\beta}}
\def\ga{{\gamma}}

\def\la{{\lambda}}

\def\bR{{\mathbb {R}}}
\def\bC{{\mathbb {C}}}

\def\bP{{\mathbb {P}}}

\def\pA{{\mathcal A}}
\def\pB{{\mathcal B}}

\def\pQ{{\mathcal Q}}

\def\pO{{\mathcal O}}
\def\pH{{\mathcal H}}

\def\tr{\mbox{\rm tr\,}}

\def\GL{{\mbox{\rm GL}}}
\def\SL{{\mbox{\rm SL}}}
\def\SO{{\mbox{\rm SO}}}
\def\Ort{{\mbox{\rm O}}}

\def\U{{\mbox{\rm U}}}

\def\rank{\mbox{\rm{rank\,}}}
\def\MatR{M_{m,n}(\bR)}
\def\SLloc{\rm SL_{loc}}
\def\GLloc{\rm GL_{loc}}

\renewcommand{\subjclassname}{\textup{2000} Mathematics Subject
Classification} 

\begin{document}

\title[Pure States of Four Qubits]
{Normal Forms and Tensor Ranks of Pure States of Four Qubits}

\author[ O. Chterental and D.\v{Z}. \DJ okovi\'{c}]
{Oleg Chterental and Dragomir \v{Z}. \DJ okovi\'{c}}

\address{Department of Pure Mathematics, University of Waterloo,
Waterloo, Ontario, N2L 3G1, Canada}

\email{ochteren@uwaterloo.ca} \email{djokovic@uwaterloo.ca}

\thanks{
The first author was supported by an NSERC Undergraduate Student Research Award, 
and the
second by  NSERC Grant A-5285}


\date{}

\maketitle \subjclassname { 81P68, 15A69}

\begin{abstract}
We examine the SLOCC classification of the non-normalized pure states of four 
qubits obtained by F. Verstraete et al. in \cite{VDDV}. The rigorous proofs of 
their basic results are provided and necessary corrections implemented. We use 
Invariant Theory to solve the problem of equivalence of non-normalized pure 
states under SLOCC transformations of determinant $1$ and qubit permutations. As 
a byproduct, we produce a new set of generators for the invariants of the Weyl 
group of type $F_4$. We complete the determination of the tensor ranks of 
four-qubit pure states initiated by J.-L. Brylinski \cite{B}. As a result we 
obtain a simple algorithm for computing these ranks. We obtain also a very 
simple classification of states of rank $\leq 3$. 
\end{abstract}

\section{Introduction}

We use the methods of Linear Algebra and Invariant Theory to study the problem 
of classification of pure quantum states of four qubits. Although we use the 
terminology common to Quantum Physics, we do not assume the reader is familiar 
with it, and we shall provide necessary definitions or references. We do not 
need a precise definition of qubits. It suffices to say that a qubit is a 
mathematical model for the quantum analog of an ordinary computer bit. A basic 
ingredient of this model is a $2$--dimensional complex Hilbert space (see 
\cite{AP,JP}).

We shall work with four qubits. The Hilbert space of the $k$-th qubit will be 
denoted by $\pH_k = \bC^2$ with an orthonormal basis $\{e_0,e_1\}$. The Hilbert 
space for the quantum system consisting of four qubits is the tensor product 
\[\pH = \pH_1 \otimes \pH_2 \otimes \pH_3 \otimes \pH_4 .\] Occasionally we 
shall use Dirac's bra-ket notation in abbreviated form, e.g., \[ |ijkl\rangle = 
e_i \otimes e_j \otimes e_k \otimes e_l. \] If $A_k$ is an invertible linear 
operator on $\pH_k$, then we refer to $A_1 \otimes A_2 \otimes A_3 \otimes A_4$ 
as an {\em invertible} SLOCC {\em operation} (reversible stochastic local 
quantum operations assisted by classical communication).

A normalized pure state is a unit vector $\psi \in \pH$ up to a phase factor. 
However we shall work mostly with non-normalized pure states, i.e., mostly with 
nonzero vectors of $\pH$ and we refer to them simply as {\em pure states}.

The classification of pure states of three qubits is now well-known for both the 
group of SLOCC operations \cite{DVC} and the group of local unitary 
transformations \cite{B}. The SLOCC classification of the pure states of four 
qubits was obtained by Verstraete et al. in \cite{VDDV}. However their list has 
an error which has not been noticed so far: the family $L_{ab_3}$ is equivalent 
to a subfamily of $L_{abc_2}$. See Remark \ref{missedfamily} for more detailed 
comments. The need to redo this classification on a more rigorous basis is also 
shared by some physicists \cite{LLSS, AM}.

The study of the tensor ranks of four-qubit pure states was initiated in a 
recent paper of J.-L. Brylinski \cite{B}, where he proposed that these ranks can 
be used as an algebraic measure of entanglement. We recall that the tensor rank 
of a pure state $\psi \in \pH$ is defined as the least number $r$ of product 
states whose sum is $\psi$. By a \mbox{\em product state} we mean a 
(non-normalized) pure state of the form $v_1 \otimes v_2 \otimes v_3 \otimes 
v_4$. It is worthy of mention that the problem of calculating tensor ranks may 
be potentially very challenging. The case of tensor products of three vector 
spaces, of arbitrary dimensions, is relevant to the theory of algebraic 
complexity in Computer Science and we refer the reader to \cite{BCS} for the 
exploration of this topic.

Our objective in this paper is threefold. First of all we shall reprove Theorems 
1 and 2 of \cite{VDDV} and at the same time improve and correct their 
formulations. Second, we give a simple method to test whether two semisimple 
states (see section \ref{m=n=4} for the definition) are equivalent under SLOCC 
operations of determinant one and qubit permutations. The case of arbitrary pure 
states is also discussed. Third, we shall present an algorithm for computing 
tensor ranks for arbitrary pure states of four qubits. Along the way we shall 
give a different and simple classification of tensors of rank $\leq 3$, we shall 
determine the Zariski closure of the tensors of rank $\leq 2$ (a question left 
open in \cite{B}), and we shall construct a nice set of generators for the 
algebra of polynomial invariants of the Weyl group of type $F_4$ (see Appendix B).

There are several groups operating on $\pH$ that are important for this paper. 
The most important ones are $\SLloc=\SL_{2} \times \SL_{2} \times \SL_{2} \times 
\SL_{2}$ and $\SLloc^*=\SLloc \cdot \rm{Sym}_4$, where $\SL_2=\SL_2(\bC)$ is the 
special linear group in two dimensions and $\rm Sym_4$ is the symmetric group on 
four symbols which acts by permuting the four qubits. Similarly define $\GLloc$ 
and $\GLloc^*$ by replacing the group $\SL_2$ with $\GL_2(\bC)$. When we refer 
to $M_n$ or $M_{m,n}$ as the spaces of $n \times n$ resp. $m \times n$ matrices 
it is assumed they are over $\bC$, otherwise we will explicitly write 
$M_n(\bR)$, etc.

Since we are interested in classifying the orbits of $\SLloc^*$, we shall need 
some basic facts about the polynomial invariants of this group. We denote by 
$\pA$ resp. $\pA^*$ the algebra of complex analytic polynomial functions on 
$\pH$ which are invariant under the action of $\SLloc$ resp. $\SLloc^*$. It is 
well known that $\pA$ is a polynomial algebra in four variables. Its generators 
are algebraically independent homogeneous polynomials $H$, $L$, $M$ and $D$ of 
degree 2,4,4 and 6 respectively. See the paper \cite{LT} where these generators 
are explicitly constructed. The quadratic invariant (first constructed by 
Cayley, see \cite{LT}) has the following simple expression 
\[H(\psi)=\sum_{i,j,k=0}^{1} (-1)^{i+j+k} \psi_{i,j,k,0} \psi_{1-i,1-j,1-k,1}.\] 
The quartic invariants $L$ and $M$ are defined in section \ref{m=n=4} and the 
sextic invariant $D$ in section \ref{criterion}.

In section \ref{complexSVD} we prove the basic result, Theorem 
$\ref{indecompreps}$, which gives the classification of indecomposable 
orthogonal representations of a very simple quiver $\pQ$ (see Figure 
$\ref{quiv}$). The matrix version of this result is stated in Theorem 
$\ref{matrixver}$. One can view this theorem as a complex analog of the real SVD 
decomposition. The paper \cite{VDDV} contains such a theorem, for the case of 
square matrices only, but the uniqueness assertion for their proposed normal 
form is not valid.

In section \ref{m=n=4} we recall some basic facts about $\pA$. We also recall 
some important facts from the theory of infinitesimal complex semisimple 
symmetric spaces and introduce the notion of semisimple and nilpotent states 
$\psi \in \pH$. The main results are stated in Theorems $\ref{classify I}$ and 
$\ref{classify II}$. They provide the classification of $\Ort_4 \times 
\Ort_4$--orbits in $M_4$ and $\SLloc^*$--orbits in $\pH$, respectively. The 
representatives of the $\Ort_4 \times \Ort_4$--orbits, organized in 17 families, 
are given in Table $\ref{orbitreps}$. Nine of the families are selected to 
obtain a set of representatives of the $\SLloc^*$--orbits in $\pH$. The explicit 
expressions for these families are listed in Table \ref{psiexpr} in appendix A.

In section \ref{criterion} we show that $\pA^*$ is also a polynomial algebra in 
four variables and exhibit its generators $H$, $\Gamma$, $\Sigma$, $\Pi$. These 
generators have degrees $2$, $6$, $8$, and $12$ and appear in a paper of 
Schl\"{a}fli in 1852. We then show that these generators can be used to separate 
semisimple orbits.

The tensors $\psi \in \pH$ of $\rank \leq 3$ can be described by elementary 
means. This is accomplished in section \ref{lowrank}. The results obtained here 
are used in an essential way in the development of our rank algorithm.

In section \ref{tensorranks} we examine the nine families giving representatives 
of $\SLloc^*$--orbits in $\pH$. For each of them we compute the tensor ranks of 
all states $\psi$ in the family. Finally, in section \ref{algo} we present a 
simple algorithm which computes the tensor rank of arbitrary pure states $\psi 
\in \pH$.

In section \ref{conc} we summarize our results and make some comments on the 
problem of equivalence of two states under the group of local unitary 
operations. From the $\SLloc^*$--classification of the pure states of four 
qubits it is easy to derive the $\GLloc^*$--classification. For a different 
approach to the $\GLloc$ and $\GLloc^*$--classifications see the recent posting 
\cite{LLSS} on the arXiv.

We thank A. Osterloh for his comments regarding the examples in
section \ref{criterion}.

\section{A complex analog of the real Singular Value 
Decomposition}\label{complexSVD}

Our classification of $\SLloc^*$--classes of pure states of four qubits is based 
on a result of Linear Algebra which we discuss in this section. 

Consider $\MatR$ where we assume $m \leq n$. By $A^T$ we denote the transpose of 
a matrix $A$. If we consider the usual action of the real orthogonal groups 
$\Ort(m)$ and $\Ort(n)$ on $\MatR$, i.e,
\begin{equation} (S_1,S_2) \cdot A = S_1 A S_2^{-1} \end{equation} where $S_1 
\in \Ort(m)$, $S_2 \in \Ort(n)$ and $A \in  M_{m,n}(\bR)$, then the orbits for 
this action are classified by diagonal matrices $\Sigma \in \MatR$ with 
non-increasing and nonnegative diagonal entries (the singular value 
decomposition theorem).

There is a complex version of this theorem where $\MatR$ is replaced by 
$M_{m,n}$, the complex $m \times n$ matrices, and $\Ort(m)$ and $\Ort(n)$ 
replaced by the unitary groups $\U(m)$ and $\U(n)$. We are interested in another 
complex version of the above theorem where instead of $\U(m)$ and $\U(n)$ we use 
the complex orthogonal groups $\Ort_m=\Ort_m(\bC)$ and $\Ort_n=\Ort_n(\bC)$. We 
recall that $\Ort_n$ is the subgroup of $\GL_n(\bC)$ consisting of all complex 
matrices $X$ such that $X^TX=I_n$. In the case $m=n$, such a theorem appears in 
the recent paper \cite{VDDV}. As the proof presented there is not completely 
clear, and the statement of the theorem we feel can be improved upon, we shall 
offer a different approach in this section.

It is convenient to use the language of quivers. We just need one very simple 
quiver $\pQ$ which has two vertices, say 1 and 2, and a single directed edge 
from 1 to 2, as in Figure \ref{quiv}.\\
\begin{figure}[h]\caption{The quiver $\pQ$}\label{quiv}
\begin{center}
\begin{picture}(65,40)(0,0)
\put(0,25){\circle{5}}
\put(55,25){\circle{5}}
\put(3,25){\vector(1,0){50}}
\put(0,10){$1$}
\put(50,10){$2$}
\end{picture}
\end{center}
\end{figure} \\
We are interested only in orthogonal representations of this quiver. For 
simplicity we shall refer to them simply as representations.

\begin{definition} \label{rep}
A representation of the above quiver $\pQ$ is a 5-tuple 
$(V_1,V_2,\phi_1,\phi_2,A)$ where $V_1$ and $V_2$ are finite dimensional complex 
vector spaces equipped with nondegenerate symmetric bilinear forms $\phi_1$ and 
$\phi_2$ respectively, and $A$ is a linear map $V_1 \rightarrow V_2$. We think 
of $V_1$ and $V_2$ as being attached to vertices 1 and 2 respectively while $A$ 
is attached to the directed edge.
\end{definition}

\begin{example} \label{basic}
The most basic example of such a representation is given by $V_1=\bC^n$, 
$V_2=\bC^m$, where $\phi_1$ and $\phi_2$ are the usual dot products on $\bC^n$ 
and $\bC^m$, and the linear transformation $A:\bC^n\rightarrow\bC^m$ is 
identified with an $m\times n$ complex matrix. Henceforth whenever we refer to 
$\bC^k$ as a vector space we will assume that the bilinear form is the usual dot 
product.
\end{example}

Let us recall some basic facts about representations of quivers in our setting. 
We start with the definition of a homomorphism of orthogonal representations.
\begin{definition} \label{homomorphism}
A homomorphism \begin{equation}S:(V_1,V_2,\phi_1,\phi_2,A) \rightarrow 
(V_1',V_2',\phi_1',\phi_2',A')\end{equation} is a pair of linear maps $S_1:V_1 
\rightarrow V_1'$ and $S_2:V_2 \rightarrow V_2'$ such that 
\begin{equation}\phi_1'(S_1 v_1, S_1 w_1)=\phi_1(v_1, w_1), \forall v_1,w_1 \in 
V_1\end{equation}
\begin{equation}\phi_2'(S_2 v_2, S_2 w_2)=\phi_2(v_2, w_2), \forall v_2,w_2 \in 
V_2\end{equation}
and $S_2 \circ A = A' \circ S_1$.
\end{definition}
If $S_1$ and $S_2$ are isomorphisms we say that $S$ is an isomorphism. If there 
exists an isomorphism between two representations, then we say that the 
representations are \mbox{\em isomorphic} or \mbox{\em equivalent}.
For instance two representations
\begin{equation} A:\bC^n \rightarrow \bC^m ;\quad B:\bC^n \rightarrow 
\bC^m\end{equation}
are isomorphic if and only if there exist $S_1 \in \Ort_n$ and $S_2 \in \Ort_m$ 
such that $S_2 A = B S_1$ i.e., $B=S_2 A S_1^{-1}$. This means that two matrices 
$A,B \in  M_{m,n}$ belong to the same orbit of $\Ort_m \times \Ort_n$, acting on 
 $M_{m,n}$ in the usual way, if and only if the two representations above are 
isomorphic. Clearly, every representation of $\pQ$ is isomorphic to one of the 
type given in example \ref{basic}.

One defines the direct sum of representations in the obvious way. A 
representation $(V_1,V_2,\phi_1,\phi_2,A)$ is nonzero if $V_1$ or $V_2$ is a 
nonzero space. A nonzero representation is said to be \mbox{\em indecomposable} 
if it is not a direct sum of two nonzero representations. The Krull--Schmidt 
theorem is valid, i.e., every representation decomposes as a direct sum of 
indecomposable ones and these indecomposable summands are unique up to 
permutation and isomorphism.

Hence the classification of all representations of our quiver $\pQ$ reduces to 
the description of its indecomposable representations (up to isomorphism).

The three simplest nonzero representations are the following.

\begin{example} \label{ex1}
The representation $A:0 \rightarrow \bC^1$ is indecomposable. The matrix of the 
linear transformation $A$ is the unique $1 \times 0$ matrix.
\end{example}

\begin{example} \label{ex2}
The representation $A:\bC^1 \rightarrow 0$ is indecomposable.  Its matrix is the 
unique $0 \times 1$ matrix.
\end{example}

\begin{example} \label{ex3}
Every representation $A:\bC^1 \rightarrow \bC^1$ is given by multiplication by a 
fixed $\al \in \bC$.  Its matrix is the $1 \times 1$ matrix $[\al]$. If $\al=0$ 
this representation is isomorphic to the direct sum of the representations from 
examples \ref{ex1} and \ref{ex2}. If $\al \neq 0$, the representation is 
indecomposable.
\end{example}

The $n \times n$ \mbox{\em symmetrized Jordan block} $J_{n}^{\times}(\al)$ is 
the sum of the band matrix having $\al$'s on the diagonal and $1$'s on the sub 
and super-diagonal and the matrix with $i$'s on the opposite super-diagonal and 
$-i$'s on the opposite sub-diagonal. For example
\[J_{5}^{\times}(\al)=\begin{bmatrix}
\al & 1 &0&i&0 \\
1& \al &1+i&0&-i\\
0&1+i& \al &1-i&0\\
i&0&1-i& \al &1\\
0&-i&0&1& \al\\
\end{bmatrix}.\]

We point out that if $n$ is odd, then the representation of $\pQ$ given by 
$J_{n}^{\times}(0)$ is decomposable. We have already seen this in Example 
\ref{ex3} when $n=1$. As a less trivial example consider
\[J_{3}^{\times}(0)=\begin{bmatrix}
0&1+i&0 \\
1+i&0&1-i \\
0&1-i&0 \\
\end{bmatrix}.
\]
By permuting rows and columns, we obtain
\[\left[ \begin{array}{c|cc}1+i&0&0 \\ 1-i&0&0 \\ \hline 0&1+i&1-i 
\\\end{array}\right].\]
Hence this representation of $\pQ$ is the direct sum of the representations 
given by the two nonzero blocks, which are indecomposable. 

\begin{remark}\label{blocks}
These blocks can be replaced with $\begin{bmatrix}1&i \\ \end{bmatrix}^T$ and 
$\begin{bmatrix}1&i \\ \end{bmatrix}$ respectively. Indeed, we have 
$\begin{bmatrix}1+i&1-i \\ \end{bmatrix}=\begin{bmatrix}1&i \\ \end{bmatrix}P$ 
where \[P=\frac{1}{4}\begin{bmatrix}3+i&3-i \\ 3-i&-3-i \\ \end{bmatrix}\] is 
orthogonal. We may simplify the symmetrized Jordan blocks as well. Note that for 
example $J_3^{\times}(\al)$ is similar to \[ \begin{bmatrix} \al&1&0 \\ 1&\al&i 
\\ 0&i&\al \end{bmatrix} \] and since they are symmetric they are orthogonally 
similar. Hence, this matrix is in the same $\Ort_3 \times \Ort_3$--orbit as 
$J_3^{\times}(\al)$.

The blocks $J_n^{\times}(\al)$ can be replaced by another kind of symmetrized 
Jordan blocks which consist of 3--diagonal symmetric matrices. For the 
description of these blocks see the recent paper \cite{DZ}.
\end{remark}

 We shall use a single matrix, as described in Example \ref{basic}, to denote a 
representation of the quiver $\pQ$. With this in mind we can now state the 
following important classification theorem.
\begin{theorem} \label{indecompreps}
The representatives of the isomorphism classes of indecomposable (orthogonal) 
representations of the quiver $\pQ$ are given by the following matrices:
\begin{enumerate}
\item $J_{n}^{\times}(\al)$ for $n \geq 1$ where $\al \neq 0$ if n is odd. The 
two values $\pm \al$ give the same isomorphism class.
\item The $(m+1)\times m$ matrix, $m \geq 0$, formed by using even index columns 
and odd index rows of $J_{2m+1}^{\times}(0)$.
\item The transpose of the previous indecomposable.
\end{enumerate}
\end{theorem}
\begin{proof}
Let us first show that the representations $A$ given in ($1$-$3$) are indeed 
indecomposable. Note that if the representations $A$ and $B$ of $\pQ$ are 
isomorphic, then the matrices $A^TA$ and $B^TB$ are similar. Consequently, the 
number of indecomposable direct summands of $A$ is at most equal to the number 
of Jordan blocks of $A^TA$.

In case $(1)$, $A=J_{n}^{\times}(\al)$. If $\al \neq 0$ then $A^TA=A^2$ has just 
one Jordan block and so the representation given by $A$ must be indecomposable. 
If $\al=0$ then $n=2m$ is even and $A^2$ is similar to $J_m(0) \oplus J_m(0)$. 
We leave to the reader to show that $A$ must be indecomposable.

The cases $(2)$ and $(3)$ can be handled together. Let $A=J_{n}^{\times}(0)$ 
with $n=2m+1$ odd. Then $A^TA=A^2$ is similar to $J_m(0) \oplus J_{m+1}(0)$ and 
so the representation $A$ has at most two indecomposable direct summands. But we 
have seen in the discussion preceding this theorem that it indeed is a direct 
sum of two representations. Hence these summands must be indecomposable.

We show next that every indecomposable representation of $\pQ$ is isomorphic to 
one of the representations ($1$-$3$).

Let $A:V \rightarrow W$ be an indecomposable representation where $V=\bC^n$ and 
$W=\bC^m$. We have that $A^TA$ and $AA^T$ are linear operators on $V$ and $W$ 
respectively. Let us apply the Fitting decomposition \begin{equation} V= V_0 
\oplus V_1 , \quad W=W_0 \oplus W_1,\end{equation}where $V_0$ and $V_1$ are 
$A^TA$-invariant subspaces, $A^TA$ is nilpotent on $V_0$ and invertible on 
$V_1$, and similar properties hold for $W_0$, $W_1$ and $AA^T$. Then it is easy 
to show that $V_0 \perp V_1$, $W_0 \perp W_1$, and $A(V_i) \subseteq W_i$ and 
$A^T(W_i) \subseteq V_i$ for $i=0,1$.

This means that the representation $A:V \rightarrow W$ is the direct sum of the 
representations $A_i:V_i \rightarrow W_i$ where $A_i$ is the restriction of $A$ 
for $i=0,1$. As our representation is assumed to be indecomposable, we have 
$V_0=W_0=0$ or $V_1=W_1=0$.

\mbox{\em Case 1}: $V_0=W_0=0$. Then $m=n$ and $A$ and $A^T$ are ismorphisms. By 
\cite{K,CH} $A$ is a product of an orthogonal matrix and a symmetric one. A 
symmetric matrix is orthogonally similar to the direct sum of symmetrized Jordan 
blocks \cite{G}. Consequently, we can write $A=PBQ$ where $P,Q\in \Ort_n$ and 
$B$ is the direct sum of symmetrized Jordan blocks. There is only one block, 
i.e, $B=J_n^{\times}(\al)$ by the indecomposability assumption. As $B$ is 
invertible, we have $\al \neq 0$. It is also required to show that two 
symmetrized Jordan blocks $J_n^{\times}(\al)$ and $J_n^{\times}(\be)$ are in the 
same isomorphism class if and only if $\al=\pm\be$. This can be seen as follows. 
If $J_n^{\times}(\al)$ and $J_n^{\times}(\be)$ give isomorphic representations 
then \[J_n^{\times}(\be)=PJ_n^{\times}(\al)Q \] for some $P,Q \in \Ort_n$. Then 
\[J_n^{\times}(\be)^2=J_n^{\times}(\be)^T J_n^{\times}(\be)=Q^T 
J_n^{\times}(\al)^2 Q =Q^{-1} J_n^{\times}(\al)^2 Q\] so $J_n^{\times}(\al)^2$ 
and $J_n^{\times}(\be)^2$ are similar and thus they have the same eigenvalues 
hence $\be^2=\al^2$. Conversely, note that $J_n^{\times}(0)$ and 
$-J_n^{\times}(0)$ are similar and symmetric so they are orthogonally similar. 
So there exists $P \in \Ort_n$ such that 
\[PJ_n^{\times}(0)P^{-1}=-J_n^{\times}(0).\] Adding $\al I_n$ to both sides we 
have \[PJ_n^{\times}(\al)P^{-1}=\al I_n 
-J_n^{\times}(0)=-J_{n}^{\times}(-\al).\] As $-I_n \in \Ort_n$ we see that 
$J_n^{\times}(\al)$ and $J_n^{\times}(-\al)$ are in the same $\Ort_n \times 
\Ort_n$--orbit, and so they give equivalent orthogonal representations of $\pQ$.

\mbox{\em Case 2}: $V_1=W_1=0$. In this case the matrices $A^TA$ and $AA^T$ are 
nilpotent. This case occurs naturally in the theory of (infinitesimal) 
semisimple complex symmetric spaces. We omit the proof and refer the reader to 
\cite{O, DLS}.
\end{proof}

\begin{remark}
The classification of indecomposables in the above theorem can be deduced from 
the general results on representations of symmetric quivers.  We refer the 
reader to the recent paper of Derksen and Weyman \cite{DW}, where this new type 
of quiver is introduced and their representations (including the orthogonal and 
symplectic ones) are studied. In order to apply their results, our quiver has to 
be modified by adding an additional directed edge from the second to the first 
vertex. The involution $\sigma$, required by the definition of symmetric 
quivers, fixes the vertices and interchanges the two arrows.
\end{remark}

We can reformulate Theorem \ref{indecompreps} in terms of matrices.
\begin{theorem}\label{matrixver}
Let $\Ort_m \times \Ort_n$ act on $M_{m,n}$ in the usual way. Then the 
block-diagonal matrices
\[ A = A_1 \oplus A_2 \oplus \cdot \cdot \cdot \oplus A_k \in M_{m,n},\]
where each $A_i$ is one of the matrices listed in Theorem $\ref{indecompreps}$, 
are representatives of $\Ort_m \times \Ort_n$--orbits. These representatives are 
unique up to permutation of the $A_i$'s and sign changes mentioned in that 
theorem.
\end{theorem}

\begin{remark}\label{compare}
This theorem should be compared with \cite[Theorem 1]{VDDV}. The authors 
consider only the square case $m=n$. Contrary to their claim, the canonical 
forms given there are not unique up to permutation of the diagonal blocks 
because some of their blocks have the shape \[\begin{bmatrix}0&R_1\\ R_2&0 \\ 
\end{bmatrix}\] and are made up of two of our rectangular blocks, one of size 
$(p+1) \times p$ and the other $q \times (q+1)$. In the formulation given in 
Theorem \ref{matrixver}, there are more possible ways of combining such blocks, 
which leads to non-uniqueness. They also failed to mention that 
$J_n^{\times}(\al)$ and $J_n^{\times}(-\al)$ belong to the same $\Ort_n \times 
\Ort_n$--orbit.
\end{remark}

\section{Classification of $\Ort_4 \times \Ort_4$--orbits in $M_4$ and 
$\SLloc^*$--orbits in $\pH$}\label{m=n=4}

Our main objective here is to apply Theorem \ref{matrixver} to the problem of 
$\SLloc^*$--classification of pure states of $\pH$. We start with some 
preliminary results, mostly well known. We have mentioned that a product state 
$\psi$ is of the form $v_1 \otimes v_2 \otimes v_3 \otimes v_4$. We say a state 
$\psi$ is \mbox{\em factorizable} if, after a permutation of qubits, it can be 
written as the product of two tensors $\psi=\psi_1 \otimes \psi_2$. As in 
\cite{VDDV}, for a tensor \[\displaystyle\psi=\sum_{i,j,k,l=0}^{1} \psi_{ijkl} 
|ijkl\rangle \] we define the $4 \times 4$ matrix $\tilde{\psi}$ by using the 
pairs $ij$ as the row index and the pairs $kl$ as the column index (we order 
these pairs as 00, 01, 10, 11). By permuting cyclically the indices $jkl$ we 
obtain two more such matrices $\tilde{\psi}'$ and $\tilde{\psi}''$. As in 
\cite{LT}, we denote their determinants by \[L=\Delta_{1234}, \quad 
M=\Delta_{1342}, \quad N=\Delta_{1423},\] respectively. It is easy to verify 
that $L+M+N=0$.

Let $S_k$ be the set of tensors with rank less than or equal to $k$. 
Surprisingly, it may happen that $S_k$ is not Zariski closed. We shall denote 
its Zariski closure by $\bar{S_k}$. We need the following result proved by 
Brylinski \cite{B}.

\begin{proposition}\label{rankfour}
The maximum rank of a tensor $\psi \in \pH$ is $4$. The affine variety 
$\bar{S_3}$ is irreducible and is defined by the equations $L=M=0$. Hence 
$\bar{S_3}$ has dimension $14$.
\end{proposition}

We prove the analogous result for $\bar{S_2}$, which was alluded to in \cite{B}.
\begin{proposition}
The affine variety  $\bar{S_2} \subseteq \pH$ is irreducible of dimension $10$. 
Its ideal is exactly the ideal generated by the forty-eight $3 \times3$ minors 
of the matrices $\tilde{\psi}$, $\tilde{\psi}'$ and $\tilde{\psi}''$.
\end{proposition}
\begin{proof}
Let $I$ be the ideal generated by the forty-eight minors and $W \subseteq \pH$ 
its zero set. We used Singular \cite{Si} to verify that $I$ is a prime ideal and 
that $\mbox{\rm dim}\ W=10$.
On the other hand a simple computation (using Maple \cite{Ma}) shows that the 
$\GLloc$--orbit $\pO$ of $|0000\rangle + |1111\rangle$ also has dimension 10. 
Since $\pO \subseteq S_2 \subseteq W$, we have $\bar{\pO} \subseteq \bar{S_2} 
\subseteq W$. As $\pO$ is dense in $W$, the proposition is proved.
\end{proof}

Consider the action of $\SL_4 \times \SL_4$ on $M_4$ given by: \[ 
((P,Q),R)\rightarrow PRQ^T. \] The image of $\SL_4 \times \SL_4$ under this 
representation is contained in $\SL(M_4)$, the special linear group of the space 
$M_4$. This image is usually written as $\SL_4 \otimes \SL_4$, which means that 
we have two copies of $\SL_4$ with their centers identified (glued together).

The action of $\SLloc$ on $\pH$ gives rise to an action on $M_4$ via the map 
$\psi \rightarrow \tilde{\psi}$. Explicitly, this action is given by 
\[((A_1,A_2,A_3,A_4),\tilde{\psi})\rightarrow (A_1 \otimes A_2) \tilde{\psi} 
(A_3 \otimes A_4)^T, \] where $A_1 \otimes A_2$ is the usual tensor product of 
matrices.

The image of the first two factors $\SL_2$ of $\SLloc$ under the action on $M_4$ 
is contained in the first factor of $\SL_4 \otimes \SL_4$. It is well known that 
this image is isomorphic to $\SL_2 \otimes \SL_2 \cong \SO_4$ but is different 
from $\SO_4$. We need to conjugate this image to obtain $\SO_4$. Clearly, the 
matrix which performs this conjugation is not unique. For that purpose we use 
the unitary matrix
\[T=\frac{1}{\sqrt{2}}\begin{bmatrix}
1&0&0&1 \\
0&i&i&0 \\
0&-1&1&0 \\
i&0&0&-i \\
\end{bmatrix},\] which we borrow from \cite{VDDV}. A slightly different such 
matrix $Q$ is given in Makhlin's paper \cite{M}. Now define 
$R=R_{\psi}=T\tilde{\psi}T^{\dag}$, where the superscript $^\dag$ indicates the 
hermitian transpose. It is assumed that the particular $\psi$ is obvious from 
the context when it is not written in the subscript. Finally we set
\[\tilde{R}=\tilde{R}_{\psi}=\begin{bmatrix}
0 & R \\
-R^T & 0 \\
\end{bmatrix}.\]

If $A_k \in \SL_2$ then $|\phi\rangle=A_1 \otimes A_2 \otimes A_3 \otimes A_4 
|\psi\rangle$ corresponds to $R_{\phi}=P_1 R_{\psi} P_2$ where $P_1, P_2 \in 
\SO_4$ are given by $P_1 = T(A_1 \otimes A_2)T^\dag$ and $P_2 = T(A_3 \otimes 
A_4)^TT^\dag$. Hence there is a 1-to-1 correspondence between the 
$\SLloc$--orbits in $\pH$ and $\SO_4 \times \SO_4$--orbits in $M_4$.

For the following facts the reader can consult chapter 38 of \cite{TY}, and in 
particular Proposition 38.6.8. The $\SLloc$--orbit of $\psi$ is closed (in the 
Zariski topology) iff the $\SO_4 \times \SO_4$--orbit of $R_{\psi}$ is closed. 
It is well known that this is the case iff the matrix $\tilde{R}_{\psi}$ is 
semisimple (i.e. diagonalizable). In this case we shall also say that $\psi$ is 
\mbox{\em semisimple}.
The Zariski closure of the $\SLloc$--orbit of $\psi$ contains the zero vector 
iff the same is true for the $\SO_4 \times \SO_4$--orbit of $R_{\psi}$. 
Furthermore, this is the case iff the matrix $\tilde{R}_{\psi}$ is nilpotent. In 
that case we shall also say that $\psi$ is \mbox{\em nilpotent}. A nilpotent 
$\SLloc$--orbit, say $\pO$, is conical, i.e., if $\psi \in \pO$ then also $\la 
\psi \in \pO$ for all nonzero scalars $\la \in \bC$. Hence $\pO$ is also a 
$\GLloc$--orbit.

For any $\psi \in \pH$, the characteristic polynomial of $\tilde{R}_{\psi}$ is 
given by \[t^8+2Ht^6+(H^2+2L+4M)t^4+2(HL+2D)t^2+L^2,\] where it is understood 
that $H$ is short for $H(\psi)$ etc. We will define the invariant $D$ in the 
next section. If $\psi \in \bar{S_3}$ then $L=M=0$ and we obtain 
\begin{equation}\label{charpoly}t^2(t^6+2Ht^4+H^2t^2+4D).\end{equation} The 
discriminant of the cubic $s^3+2Hs^2+H^2s+4D$ is equal to 
\begin{equation}\label{discr}16D(H^3-27D).\end{equation}

The conjugation by the diagonal matrix $I_4 \oplus (-I_4)$ induces an involutory 
automorphism $\theta$ of $\Ort_8$ and its Lie algebra $\mbox{\germ g} = 
\gso_8(\bC)$ consisting of the skew-symmetric matrices in $M_8$. Let $\mfk$ and 
$\mfp$ be the eigenspaces of $\theta$ in $\gg$ with eigenvalues $+1$ and $-1$, 
respectively. These eigenspaces consist of the matrices having the form 
\[\begin{bmatrix}\star & 0 \\ 0 & \star \end{bmatrix} \quad \mbox{\rm resp.} 
\quad \begin{bmatrix} 0& \star \\ \star & 0 \end{bmatrix},\] all blocks being of 
size 4. The space $\mfk$ is in fact a subalgebra of $\gg$, the Lie algebra of 
the subgroup $K=\Ort_4 \times \Ort_4$ of $\Ort_8$. The space $\mfp$ is a 
$K$-module with the action \[\left(\begin{bmatrix}P_1 & 0 \\ 0 & P_2 
\end{bmatrix},\begin{bmatrix} 0 & R \\ -R^T & 0 \end{bmatrix}\right) \rightarrow 
\begin{bmatrix}0 & P_1 R P_2^{-1} \\ -P_2 R^T P_1^{-1} & 0\end{bmatrix}.\] This 
is an example of an (infinitesimal) semisimple complex symmetric space. The 
following theorem is a special case of general results about such spaces 
\cite[Lemma 38.7.14]{TY}:

\begin{theorem}\label{orbit}
Let $\phi,\psi \in \pH$ be semisimple states. The invariants $H$, $L$, $M$, $D$ 
take the same values at $\phi$ and $\psi$ iff $R_{\phi}$ and $R_{\psi}$ belong 
to the same $\SO_4 \times \SO_4$--orbit, i.e., $\phi$ and $\psi$ belong to the 
same $\SLloc$--orbit.
\end{theorem}

The special case $m=n=4$ of Theorem \ref{matrixver} plays an important role in 
the sequel. We now state this special case in more detail.

\begin{theorem} \label{classify I}
The $17$ families of matrices $R$, listed in Table $\ref{orbitreps}$, classify 
the $\Ort_4 \times \Ort_4$ orbits on $M_4$ up to permutation of diagonal blocks 
of the same size and replacing the parameters $a,b,c,d$ by $\pm a$, $\pm b$, 
$\pm c$, $\pm d$ respectively.
\end{theorem}

\begin{table}[htbp]\caption{$R$--matrix representatives of $\Ort_4 \times 
\Ort_4$--orbits}\label{orbitreps}
\begin{center}
\begin{tabular*}{0.75\textwidth}{ll}
1. $\left[\begin{array}{c|c|c|c} a&&& \\ \hline &b&& \\ \hline &&c& \\ \hline 
&&&d \\ \end{array}\right]$ & 
2. $\left[\begin{array}{c|c|cc} a&&& \\ \hline &b&& \\ \hline &&c+i&1 \\ &&1&c-i 
\\ \end{array}\right]$ \\\\
3. $\left[\begin{array}{c|c|c|c} a&&& \\ \hline &b&& \\ \hline &&1& \\  &&i& 
\end{array}\right]$ &
4. $\left[\begin{array}{c|c|cc} a&&& \\ \hline &b&& \\ \hline &&1&i \\ \hline 
&&&\end{array}\right]$ \\\\
5. $\left[\begin{array}{cc|cc} a+i&1&& \\ 1&a-i&& \\ \hline &&b+i&1 \\ 
&&1&b-i\end{array}\right]$ & 
6. $\left[\begin{array}{c|ccc} a&&& \\ \hline &b&1&0 \\ &1&b&i \\ 
&0&i&b\end{array}\right]$ \\\\
7. $\left[\begin{array}{c|cc|c} a&&& \\ \hline &1&i& \\  &1+i&1-i& \\  
&-i&1&\end{array}\right]$ & 
8. $\left[\begin{array}{c|ccc} a&&& \\ \hline &1&1+i&-i \\ &i&1-i&1 \\ \hline 
&&&\end{array}\right]$ \\\\
9. $\left[\begin{array}{cccc} a&1&i&0 \\ 1&a+i&1&-i \\ i&1&a-i&1 \\ 
0&-i&1&a\end{array}\right]$ &
10. $\left[\begin{array}{cc|c|c} a+i&1&& \\ 1&a-i&& \\ \hline &&1& \\ 
&&i&\end{array}\right]$ \\\\
11. $\left[\begin{array}{cc|cc} a+i&1&& \\ 1&a-i&& \\ \hline &&1&i \\ \hline 
&&&\end{array}\right]$ & 
12. $\left[\begin{array}{ccc|c} 1&0&i& \\ 1&1+i&-i& \\ i&1-i&1& \\ 
-i&0&1&\end{array}\right]$\\\\
13. $\left[\begin{array}{cccc} 1&1&i&-i \\ 0&1+i&1-i&0 \\ i&-i&1&1 \\ \hline 
&&&\end{array}\right]$ & 
14. $\left[\begin{array}{cc|cc} 1&i&& \\ 1+i&1-i&& \\ -i&1&& \\ \hline 
&&1&i\end{array}\right]$ \\\\
\end{tabular*}
\end{center}
\end{table}

\begin{table}[htbp]
\begin{center}
\begin{tabular*}{0.75\textwidth}{ll}
15. $\left[\begin{array}{ccc|c} 1&1+i&-i& \\ i&1-i&1& \\ \hline &&&1 \\ 
&&&i\end{array}\right]$ &
16. $\left[\begin{array}{c|c|c|c} 1&&& \\ i&&& \\ \hline &1&& \\ 
&i&&\end{array}\right]$ \\\\
17. $\left[\begin{array}{cc|cc} 1&i&& \\ \hline &&1&i \\ \hline &&& \\ \hline 
&&&\end{array}\right]$ &
\end{tabular*}
\end{center}
\end{table}

It should be noted that the representatives given in Table \ref{orbitreps} may 
contain blocks different from those listed in Theorem \ref{indecompreps}; some 
of them have been simplified using Remark \ref{blocks}. Table \ref{jordanstruc} 
describes the Jordan structure of the $\tilde{R}_{\psi}$ matrices. The $1 \times 
1$ Jordan blocks are given by listing their eigenvalues. The symbol $J_2(\pm 
ic)$ indicates two $2 \times 2$ Jordan blocks with eigenvalues $ic$ and $-ic$ 
respectively, etc. The family 1 consists of semisimple elements. On the other 
hand none of the other families contains a semisimple element. The nilpotent 
$\Ort_4 \times \Ort_4$--orbits are easy to identify by using Table 
\ref{jordanstruc}: Just set all parameters (if any) to $0$ in each of the 17 
families.

Table \ref{correspondence} gives a correspondence between the families of orbits 
found in \cite{VDDV} and those that we have identified. More precisely, for each 
of the $9$ families given in \cite{VDDV} by explicit expressions we have 
determined the corresponding matrices $\tilde{R}_{\psi}$ and their Jordan 
structure as well as the corresponding $\Ort_4 \times \Ort_4$--family in our 
notation (see Table \ref{orbitreps}). The appearance of the imaginary units $i$ 
in the expressions for eigenvalues of $\tilde{R}_{\psi}$ is due to the fact that 
this matrix is skew-symmetric while the matrix $P$ used in \cite{VDDV} is 
symmetric.

\begin{remark}\label{missedfamily} Verstraete et al. \cite{VDDV} state that they 
found only 12 $\Ort_4 \times \Ort_4$--families, while we found 17. This is 
probably due to the fact that their Theorem $1$ is not correct as stated. Their 
family $L_{ab_3}$ is equivalent to the subfamily of $L_{abc_2}$ obtained by 
setting $c=a$. We believe that there are two misprints in the formula for 
$L_{ab_3}$: the two $+$ signs, in the last line of this formula, should be 
replaced by $-$ signs. After this change, the family $L_{ab_3}$ is equivalent to 
our family 6 and we have a perfect correspondence between their nine 
$\SLloc^*$--families and ours.\end{remark}

Some groups of families become one family once we examine how they behave under 
permutations of qubits. That is, an $\SLloc$--orbit from one family may be taken 
to an orbit in another family by permuting qubits. After this consideration 
there are nine different groups of families as found in \cite{VDDV}.  They are 
$\{1\}$,$\{2\}$,$\{3,4,5\}$,$\{6\}$,$\{7,8,9\}$,$\{10,11\}$,$\{12,13\}$, 
$\{14,15\}$ and $\{16,17\}$.

\begin{table}[htb]\caption{Jordan structure of 
$\tilde{R}_{\psi}$}\label{jordanstruc}
\begin{center}
\begin{tabular}{ll}
1. $\pm ia$, $\pm ib$, $\pm ic$, $\pm id$ & \quad 2. $\pm ia$, $\pm ib$, 
$J_2(\pm ic)$ \\\\

3. $\pm ia$, $\pm ib$, $0$, $J_3(0)$ & \quad 6. $\pm ia$, $J_3(\pm ib)$ \\
4. Same as 3 & \\
5. $J_2(\pm ia)$, $J_2(\pm ib)$ & \\\\

7. $\pm ia$, $0$, $J_5(0)$ & \quad 10. $J_2(\pm ia)$, $0$, $J_3(0)$ \\
8. Same as 7 & \quad 11. Same as 10 \\
9. $J_4(\pm ia)$ & \\\\

12. $0$, $J_7(0)$ & \quad 14. $J_3(0)$, $J_5(0)$ \\
13. Same as 12 & \quad 15. Same as 14 \\\\

16. $0$, $0$, $J_3(0)$, $J_3(0)$ &\\
17. Same as 16 &\\
\end{tabular}
\end{center}
\end{table}

\begin{theorem} \label{classify II}
The orbits of $\SLloc^*$ on $\pH$ are classified by the nine families 
$1$,$2$,$3$,$6$,$9$,$10$,$12$,$14$ and $16$ listed in appendix A (Table 
$\ref{psiexpr}$). Their $R$--matrices are given in Table $\ref{orbitreps}$. 
States belonging to two different families (from this list of nine) are not 
equivalent under $\SLloc^*$--operations. However, within the same family, 
different values of the parameters may give states belonging to the same 
$\SLloc^*$--orbit.
\end{theorem}
\begin{proof}
Denote by $R_i$ the $R$--matrix of the $i-$th family as given in Table 
\ref{orbitreps}. Assume that $k \in \{3,7,10,12,14,16\}$. One can easily compute 
the new $R$--matrix, $R_k'$, which results by applying the permutation 
$(1,4)(2,3)$ of the four qubits. Then it is easy to see that after multiplying 
the first row and the first column of $R_k'$ by $-1$, we obtain exactly the 
transpose of $R_k$ (if $k=3$ or $7$ this step is redundant). By inspection of 
Table \ref{orbitreps} we see that $R_k^T=R_{k+1}$. This means that the $k$-th 
and $(k+1)$-st family of $\Ort_4 \times \Ort_4$--orbits fuse into a single 
family of $\SLloc^*$--orbits.

We leave to the reader to verify that the family $5$ resp. $9$ fuses with the 
family $3$ resp. $7$ into a single $\SLloc^*$--family.
\end{proof}

Let us point out that the $\Ort_4 \times \Ort_4$--orbits may be disconnected and 
that different connected components may behave differently under qubit 
permutations. We shall illustrate this in the case of the families $14$ and 
$15$. These families are in fact single $\Ort_4 \times \Ort_4$--orbits which we 
denote as $\pO_{14}$ and $\pO_{15}$, respectively. Each of them has two 
connected components: \[ \pO_{14} = \pO_{14}^I \cup \pO_{14}^{II}, \quad 
\pO_{15} = {^I}\pO_{15} \cup {^{II}}\pO_{15}. \] These facts and some others 
that we will use are explained in \cite{DLS}. To be precise, we assume that the 
Roman superscripts I and II are chosen so that the representative of $\pO_{14}$ 
resp. $\pO_{15}$ given in Table $\ref{orbitreps}$ belongs to $\pO_{14}^I$ resp. 
$^I\pO_{15}$. The left resp. right multiplication of $R$ by an orthogonal matrix 
with determinant $-1$ has the effect of switching the first two resp. last two 
qubits. The former leaves $\pO_{14}^I$ and $\pO_{14}^{II}$ invariant and 
switches $^I\pO_{15}$ and $^{II}\pO_{15}$, the latter switches $\pO_{14}^I$ and 
$\pO_{14}^{II}$ and leaves $^I\pO_{15}$ and $^{II}\pO_{15}$ invariant. Switching 
qubits $2$ and $3$ is a new feature, not discussed in \cite{DLS}. We claim that 
in this case it has the following effect: The components $\pO_{14}^I$ and 
$^{II}\pO_{15}$ get interchanged while the components $\pO_{14}^{II}$ and 
$^I\pO_{15}$ remain invariant. To carry out this verification, one cannot rely 
on the Jordan structure of the matrices $\tilde{R}$ as they are the same for 
both $\pO_{14}$ and $\pO_{15}$. However these orbits have different 
$ab$--diagrams which makes it possible to verify the claim. For the definition 
of $ab$--diagrams see \cite{KP,O,DLS}. Since the transpositions $(1,2)$, $(2,3)$ 
and $(3,4)$ generate $\rm Sym_4$, it follows that $\pO_{14} \cup \pO_{15}$ is 
indeed a single $\SLloc^*$--orbit.

The redundancies mentioned in Theorem $\ref{classify II}$ will be addressed in 
the next section.

\begin{table}[h]\caption{Correspondence of families of 
orbits}\label{correspondence}
\begin{center}
\begin{tabular*}{0.75\textwidth}{lll}
Family in \cite{VDDV} & Jordan blocks of $\tilde{R}_{\psi}$ & Our family \\
\hline & & \\
$G_{abcd}$ & $\pm ia$, $\pm ib$, $\pm ic$, $\pm id$ & $1$ \\
$L_{abc_2}$ & $\pm ia$, $\pm ib$, $J_2(\pm ic)$ & $2$ \\
$L_{a_2 b_2}$ & $J_2(\pm ia)$, $J_2(\pm ib)$ & $5$ \\
$L_{ab_3}$ & $\pm ia$, $\pm ib$, $J_2(\pm ia)$ & ? \\
$L_{a_4}$ & $J_4(\pm ia)$ & $9$ \\
$L_{a_2 0_{3 \oplus \bar{1}}}$ & $J_2(\pm ia)$, $0$, $J_3(0)$ & $10$ \\
$L_{0_{5 \oplus \bar{3}}}$ & $J_3(0)$, $J_5(0)$ & $15$ \\
$L_{0_{7 \oplus \bar{1}}}$ & $0$, $J_7(0)$ & $13$ \\
$L_{0_{3+\bar{1}},0_{3+\bar{1}}}$ & $0$, $0$, $J_3(0)$, $J_3(0)$ & $16$ 
\end{tabular*}
\end{center}
\end{table}

\section{Criterion for $\SLloc^*$--equivalence}\label{criterion}

In this section we give a criterion for testing the equivalence of two states 
$\phi,\psi \in \pH$ under $\SLloc^*$--operations. In the case when $\phi$ and 
$\psi$ are semisimple the criterion is very easy to use: one just has to verify 
whether the four invariants $H$,$\Gamma$,$\Sigma$,$\Pi$ (the latter three to be 
defined below) take the same values on $\phi$ and $\psi$.

As mentioned in the introduction, the algebra $\pA$ of complex analytic 
polynomial functions $f:\pH \rightarrow \bC$ which are $\SLloc$--invariant, 
i.e., satisfy \[ f(g \cdot \psi) = f(\psi), \quad \forall g \in \SLloc, \forall 
\psi \in \pH, \] is isomorphic to a polynomial algebra in four generators. 
Explicit generators, as constructed in \cite{LT}, are $H$,$L$,$M$ and another 
polynomial $D$. The definition of $D$ is somewhat involved.

For $j,k \in \{0,1\}$ let \[ \Psi_{jk} = \sum_{i,l=0}^1 \psi_{ijkl} x_i y_l \] 
where $x_0$,$x_1$,$y_0$,$y_1$ are independent commuting indeterminates. The 
determinant \[ \left | \begin{matrix} \Psi_{00} & \Psi_{01} \\ \Psi_{10} & 
\Psi_{11} \end{matrix} \right | \] is a biquadratic form in the two sets of 
variables $\{x_0,x_1\}$ and $\{y_0,y_1\}$. There is a unique $3 \times 3$ matrix 
$B$ such that this form can be written as 
\[ \begin{bmatrix} x_0^2 & x_0 x_1 & x_1^2 \end{bmatrix} B \begin{bmatrix} y_0^2 
\\ y_0 y_1 \\ y_1^2 \end{bmatrix}. \] Then $D(\psi)= \det B \in \pA$ and it is 
homogeneous of degree 6. By permuting cyclically the last three indices of 
$\psi$, we obtain two more such invariants which we denote by $E$ and $F$.

One can easily verify that the four homogeneous polynomials \[ H, \quad \Gamma = 
D+E+F, \quad \Sigma=L^2+M^2+N^2, \] \[ \Pi=(L-M)(M-N)(N-L), \] are algebraically 
independent and invariant under the action of $\SLloc^*$. The degrees of these 
polynomials are $2$,$6$,$8$ and $12$, respectively. These polynomials appear in 
a work of Schl\"{a}fli in 1852, who also noticed their invariance property under 
permutations of indices \cite{S}.

Let us now examine the mentioned redundancies of Theorem \ref{classify II}. The 
most interesting case is that of family 1, the family of all semisimple orbits. 
The question we raise is the following: When are two states $\psi_{abcd}$ and 
$\psi_{a'b'c'd'}$ in the same $\SLloc^*$--orbit? (By $\psi_{abcd}$ we denote the 
state whose $R$--matrix is the first matrix in Table $\ref{orbitreps}$.)

Let $\mfa$ be the subspace of $\pH$ consisting of tensors $\psi$ with $R_\psi$ a 
diagonal matrix. If we identify $\pH$ with $\mfp$ using the map $\psi 
\rightarrow \tilde{R}_\psi$ then $\mfa$ is a maximal abelian subspace of $\mfp$ 
consisting of semisimple elements. Such a subspace is known as a Cartan subspace 
of $\mfp$. We mention that all Cartan subspaces of $\mfp$ are conjugate by 
$\SO_4 \times \SO_4$, the identity component of the subgroup $K=\Ort_4 \times 
\Ort_4$ of $\Ort_8$. 

Let $N_\mfa$ resp. $Z_\mfa$ be the subgroup of $\SLloc$ which leaves $\mfa$ 
globally resp. pointwise invariant. Define similarly the subgroups $N_\mfa^*$ 
and $Z_\mfa^*$ of $\SLloc^*$. The quotient groups $W_\mfa=N_\mfa / Z_\mfa$ and 
$W_\mfa^*=N_\mfa^* / Z_\mfa^*$ act on $\mfa$ effectively.

Let us use the diagonal entries of $R_\psi$ as coordinates in $\mfa$. It is easy 
to see that $W_\mfa$ can permute arbitrarily the coordinates $a$,$b$,$c$,$d$ and 
also replace them with $\pm a$, $\pm b$, $\pm c$, $\pm d$ provided the number of 
$"-"$ signs is even. We conclude that $W_\mfa$ has order at least $192$. On the 
other hand, $\mfa$ is a Cartan subalgebra of $\gso_8$ and $W_\mfa$ is a subgroup 
of the Weyl group of the pair $(\gso_8 , \mfa)$. Since $\gso_8$ has Cartan type 
$D_4$, this Weyl group has exactly order $192$. We conclude that $W_\mfa$ 
coincides with this Weyl group.

It is easy to check that ${\rm Sym_4} \subset N_\mfa^*$, i.e., all qubit 
permutations map $\mfa$ into itself. It is also easy to check that the 
permutation $(1,2)(3,4)$ acts trivially on $\mfa$. It follows that the Klein 
four-group $V\triangleleft {\rm Sym_4}$ also acts trivially. The transposition 
$(2,3)$ sends the point $(a,b,c,d)$ to the point \[ 
\frac{1}{2}(a+b+c+d,a+b-c-d,a-b+c-d,a-b-c+d), \] i.e., it acts as the reflection 
in the hyperplane $a=b+c+d$. By using $\rm GAP$ \cite{Ga}, one can easily check 
that $W_\mfa$ and this reflection generate a group of order $1152$. Clearly this 
is the Weyl group of type $F_4$.

\begin{lemma}\label{weyl}
$W_\mfa^*$ is the Weyl group of type $F_4$, of order $1152=2^7 \cdot 3^2$.
\end{lemma}

\begin{proof}
We have \begin{equation}\label{eq} [N_\mfa^* : Z_\mfa]=[N_\mfa^* : 
N_\mfa][N_\mfa : Z_\mfa]=[N_\mfa^* : Z_\mfa^*][Z_\mfa^* : Z_\mfa]. 
\end{equation} We have seen above that $[N_\mfa^* : Z_\mfa^*]=|W_\mfa^*| \geq 
1152$. Since the Klein four-group $V \triangleleft \rm Sym_4$ acts trivially on 
$\mfa$, we have $[Z_\mfa^* : Z_\mfa] \geq 4$. On the other hand, since 
$N_\mfa=N_\mfa^* \cap \SLloc$, we have $[N_\mfa^* : N_\mfa] \leq [\SLloc^* : 
\SLloc] = 24$. Recall also that $[N_\mfa : Z_\mfa]=192$. The equality \ref{eq} 
implies now that all these inequalities are in fact equalities. In particular, 
\[ |W_\mfa^*|=[N_\mfa^* : Z_\mfa^*]=1152. \] Since $W_\mfa$ is a finite 
irreducible reflection group of rank $4$, it must be the Weyl group of type 
$F_4$.
\end{proof}

\begin{theorem}\label{polyalgiso}
The restriction homomorphism $\rho:\pA^* \rightarrow \pB$ from the algebra 
$\pA^*$ to the algebra $\pB$ of polynomial $W_\mfa^*$ invariants on $\mfa$ is an 
isomorphism of graded algebras. The algebra $\pA^*$ is generated by the four 
homogenous algebraically independent polynomials $H$, $\Gamma$, $\Sigma$, and 
$\Pi$ of degree $2$, $6$, $8$ and $12$, respectively.
\end{theorem}

\begin{proof}
If $f \in {\rm ker} \rho$, i.e., $f\in\pA^*$ and $f$ vanishes on $\mfa$, then 
$f$ vanishes on all semisimple elements of $\pH$. But the semisimple elements 
are dense in $\pH$, and so $f \equiv 0$. This shows that $\rho$ is injective.

In view of Lemma \ref{weyl}, we can apply to $W_\mfa^*$ some well known facts 
about finite reflection groups, see for example \cite[Section 3.7]{H}. The 
algebra $\pB$ is isomorphic to a polynomial algebra in four variables and it is 
generated by four homogeneous polynomials of degree $2$, $6$, $8$ and $12$. 
Moreover any set of four homogeneous generators of $\pB$ must have these 
degrees. Now recall that the $\SLloc^*$--invariants $H$, $\Gamma$, $\Sigma$, 
$\Pi$ have exactly these degrees. Since they are algebraically independent, and 
$\rho$ is injective, their restrictions to $\mfa$ are also algebraically  
independent. As their degrees are $2$, $6$, $8$ and $12$, they must generate 
$\pB$. Hence $\rho$ is also surjective. We can now prove the following analog of 
Theorem \ref{orbit}.
\end{proof}

\begin{theorem}
Two semisimples states $\phi, \psi \in \pH$ are $\SLloc^*$--equivalent iff the 
invariants $H$,$\Gamma$,$\Sigma$ and $\Pi$ take the same values at $\phi$ and 
$\psi$. For arbitrary states $\phi,\psi \in \pH$, if at least one of the 
invariants $H$, $\Gamma$, $\Sigma$, $\Pi$ takes different values on $\phi$ and 
$\psi$, then the $\SLloc^*$--orbits of $\phi$ and $\psi$ are different.
\end{theorem}
\begin{proof}
The second assertion is obvious. For the first assertion, we need only prove 
that its condition is sufficient. Assume that the condition is satisfied. Let 
$\mfa \subset \pH$ be the Cartan subspace introduced above. Since every 
semisimple element $\psi \in \pH$ is $\SLloc$--equivalent to an element of 
$\mfa$, we may assume that $\phi, \psi \in \mfa$. By Theorem \ref{polyalgiso} 
and our hypothesis, all invariants of $W_\mfa^*$ take the same values on $\phi$ 
and $\psi$. Since $W_\mfa^*$ is a finite reflection group, we conclude that 
$\phi$ and $\psi$ are $W_\mfa^*$--equivalent. Since $W_\mfa^*=N_\mfa^* / 
Z_\mfa^*$, it follows that $\phi$ and $\psi$ are $N_\mfa^*$--equivalent. In 
particular, they are $\SLloc^*$--equivalent.
\end{proof}

We can now use this theorem to show in a straightforward manner that certain 
tensors are not $\SLloc^*$--equivalent. In some situations one may use certain 
Bell inequalities as in \cite{WYKO} to show that two states are not equivalent. 
However we feel that simply calculating the $\SLloc^*$--invariants is a more 
straightforward approach and the above theorem will be enough for the majority 
of situations.

\begin{table}[t]\caption{$\SLloc^*$--invariants of some pure 4-qubit 
states}\label{examples}
\begin{center}
\begin{tabular}{lcccc}
State & $H$ & $\Gamma$ & $\Sigma$ & $\Pi$ \\ \hline \\
$GHZ$ & $\frac{1}{2}$ & $0$ & $0$ & $0$ \\\\
$W$ & $0$ & $0$ & $0$ & $0$ \\\\
$|\phi\rangle$ & $0$ & $0$ & $\frac{1}{128}$ & $\frac{1}{2048}$ \\\\
$|\phi'\rangle$ & $0$ & $0$ & $\frac{1}{128}$ & $\frac{1}{2048}$ \\\\
$|\chi\rangle$ & $0$ & $0$ & $\frac{1}{128}$ & $\frac{-1}{2048}$\\
\end{tabular}
\end{center}
\end{table}

Let us look at some examples. The generalized $GHZ$ and $W$ states in four 
qubits are \[ \frac{1}{\sqrt{2}}(|0000\rangle + |1111\rangle) \] and 
\begin{equation}\label{Wstate} 
\frac{1}{2}(|0001\rangle+|0010\rangle+|0100\rangle+|1000\rangle) \end{equation} 
respectively. Two important states in quantum teleportation
\cite[Eq. (22) and Eq. (2)]{WYKO} are the cluster state
\[ |\phi\rangle = \frac{1}{2}(|\beta^+ 0 \beta^+ 0\rangle + 
|\beta^+ 0 \beta^- 1\rangle + |\beta^- 1 \beta^- 0\rangle + |\beta^- 1 \beta^+ 
1\rangle) \] where $|\beta^\pm\rangle = \frac{1}{\sqrt{2}}(|0\rangle \pm 
|1\rangle)$ and the state $|\chi\rangle$ given by
\begin{eqnarray*}
|\chi\rangle &=& \frac{1}{2\sqrt{2}} \left( |0000\rangle - |0011\rangle - 
|0101\rangle + |0110\rangle \right. \\
&& \left. \quad + |1001\rangle + |1010\rangle + |1100\rangle + 
|1111\rangle \right).
\end{eqnarray*}
There is also another cluster state mentioned in \cite[Eq. (4)]{BR}:
\[ |\phi ' \rangle = \frac{1}{2}(|0000\rangle + |0011\rangle + |1100\rangle - 
|1111\rangle). \]

In Table \ref{examples} we tabulate the values of $H$, $\Gamma$, $\Sigma$ and 
$\Pi$ on these five states. It is clear from this table that all five states 
belong to different $\SLloc^*$--orbits except possibly for the pair
$|\phi\rangle$ and $|\phi ' \rangle$. These two states share the same
invariants and, as both are semisimple, they indeed belong to the same
$\SLloc^*$--orbit.

We remark that the states $e^{i\pi/4}|\phi\rangle$ and $|\chi\rangle$
belong to the same $\SLloc^*$--orbit because they are both semisimple
and share the same invariants. As A. Osterloh
has pointed to us, from the quantum physics point of view,
the unit vectors $ |\phi\rangle$ and $e^{i\pi/4}|\phi\rangle$ represent
the same pure state because they differ only by a phase factor. 
On the other hand, they belong to different $\SLloc^*$--orbits.
The apparent discrepancy is explained by the fact
that we are classifying the nonzero vectors in $\pH$ rather than 
the genuine pure states (see the Introduction).

We can now sketch our procedure that one can use to decide whether two arbitrary 
pure states $\phi, \psi \in \pH$ are $\SLloc^*$--equivalent. It is understood 
that equivalence will mean $\SLloc^*$--equivalence for the rest of this section. 
Clearly if $\phi$ and $\psi$ are equivalent and $\phi$ is semisimple or 
nilpotent then $\psi$ must have the same property.

{\em Step 1.} We compute the values of $H$, $\Gamma$, $\Sigma$ and $\Pi$ at 
$\phi$ and $\psi$. If they do not agree then $\phi$ and $\psi$ are not 
equivalent. From now on we assume that they do agree. If $\phi$ and $\psi$ are 
semisimple, they must be equivalent. We shall now assume they are not 
semisimple.

{\em Step 2.} Assume $\phi$ and $\psi$ are nilpotent and compute the Jordan 
structures of $\tilde{R}_\phi$ and $\tilde{R}_\psi$. By inspecting Table 
\ref{jordanstruc}, with all eigenvalues set to 0, we see that apart from one 
case the Jordan structure of the $\tilde{R}$--matrix determines uniquely the 
$\SLloc^*$--orbit. The exceptional case is when the Jordan blocks are of size 
$1,1,3,3$. Then there are two orbits. They can be distinguished by using 
$ab$--diagrams. One of these orbits is that of the generalized $W$ state and the 
other is in family $16$ (or $17$). From now on we assume that $\phi$ and $\psi$ 
are not nilpotent.

{\em Step 3.} The families $5$ and $9$ can be distinguished from the famlies 
$3$, $4$ and $7$, $8$ respectively by the sizes of Jordan blocks of 
$\tilde{R}$--matrices. By permuting qubits in both $\phi$ and $\psi$ we can 
assume that both states belong to one of the families $2$, $5$, $6$, $9$, $10$ 
or $11$. After this reduction $\phi$ and $\psi$ are equivalent iff they have the 
same Jordan structure.

\section{Classification of tensors of rank at most three}\label{lowrank}

Here we provide some normal forms for tensors of rank 1,2 and 3 under the action 
of $\SLloc^*$ and investigate some of their properties. A rank 1 tensor is just 
a product state and so it is in the same orbit as $|0000\rangle$. 

For the rank 2 case we have a few more situations to consider. Let 
\[\begin{tabular}{lll}
$\psi$ & $=$ & $a_1 \otimes a_2 \otimes a_3 \otimes a_4$ \\
&$+$ & $b_1 \otimes b_2 \otimes b_3 \otimes b_4$
\end{tabular}\]
be a rank $2$ tensor. We may consider where linear dependencies occur amongst 
the sets $\{a_i,b_i\}$.  Since we also consider the action of $\rm{Sym}_4$ there 
are really only 3 cases as in Figure \ref{ranktwo}.
\begin{figure}\caption{Patterns for rank 2 tensors}\label{ranktwo}
\begin{center}
\begin{tabular}{lll}
(a) \quad \begin{picture}(65,16)(0,7)
\put(0,0){\circle*{2}}
\put(15,0){\circle*{2}}
\put(30,0){\circle*{2}}
\put(45,0){\circle*{2}}
\put(0,16){\circle*{2}}
\put(15,16){\circle*{2}}
\put(30,16){\circle*{2}}
\put(45,16){\circle*{2}}
\end{picture} & 
(b) \quad \begin{picture}(65,16)(0,7)
\put(0,0){\circle*{2}}
\put(15,0){\circle*{2}}
\put(30,0){\circle*{2}}
\put(45,0){\circle*{2}}
\put(0,16){\circle*{2}}
\put(15,16){\circle*{2}}
\put(30,16){\circle*{2}}
\put(45,16){\circle*{2}}
\put(0,0){\line(0,1){16}}
\end{picture} & 
(c) \quad\begin{picture}(65,16)(0,7)
\put(0,0){\circle*{2}}
\put(15,0){\circle*{2}}
\put(30,0){\circle*{2}}
\put(45,0){\circle*{2}}
\put(0,16){\circle*{2}}
\put(15,16){\circle*{2}}
\put(30,16){\circle*{2}}
\put(45,16){\circle*{2}}
\put(0,0){\line(0,1){16}}
\put(15,0){\line(0,1){16}}
\end{picture}
\end{tabular}
\end{center}
\end{figure} A line connecting two points in the $i$-th column means those 
corresponding two vectors are scalar multiples of each other. In case $(a)$ let 
$g_i \in \GL_2$ be such that $g_i(a_i)=\nu_i e_0$ and $g_i(b_i)=\nu_i e_1$ where 
$\nu_i$ is chosen so that $g_i \in \SL_2$, for $i=1,2,3,4$. Hence the tensor 
reduces to $\al (|0000\rangle + |1111\rangle)$, where $\al=\nu_1 \nu_2 \nu_3 
\nu_4$. If $\{a_i,b_i\}$ is linearly dependent then we may set $\nu_i=1$ since 
we are free to choose how the transformation acts on some other vector that 
creates a basis. Let us also describe how to handle $(b)$. Since $\{a_1,b_1\}$ 
is linearly dependent, we may assume $b_1=a_1$. Hence the tensor is in the same 
orbit as $\al |0\rangle \otimes (|000\rangle + |111\rangle)$. Similarly $(c)$ is 
in the same orbit as $\al |00\rangle \otimes (|00\rangle+|11\rangle)$. Note that 
conversely any tensor in one of these forms is of rank 2.

\begin{proposition}\label{ranktwonormal}
A rank $2$ tensor $\psi \in \pH$ is $\SLloc^*$--equivalent to one of the 
following:\\\\
$(a)$ \quad $\al (|0000\rangle + |1111\rangle)$, $\al \neq 0$, \\
$(b)$ \quad $|0\rangle \otimes (|000\rangle + |111\rangle)$, \\
$(c)$ \quad $|00\rangle \otimes (|00\rangle+|11\rangle)$. \\
In case $(a)$ $\psi$ is semisimple and non-factorizable, while in cases $(b)$ 
and $(c)$ it is nilpotent and factorizable.
\end{proposition}
\begin{proof}
The first assertion follows from the above discussion. We can assume that 
$\al=1$ in cases $(b)$ and $(c)$ since they are nilpotent orbits and so 
$\SLloc^*$ and $\GLloc^*$--orbits coincide. The second assertion is easy to 
verify.
\end{proof}

The case of a rank 3 tensor is not as easy to breakdown. The complications arise 
because now we have 3 vectors $a_i, b_i, c_i$ being mapped under a 
$\SL_2$--transformation but can only control where 2 of them are mapped to in 
most cases. Let
\[\begin{tabular}{lll}
$\psi$ & $=$ & $a_1 \otimes a_2 \otimes a_3 \otimes a_4$ \\
&$+$ & $b_1 \otimes b_2 \otimes b_3 \otimes b_4$ \\
&$+$ & $c_1 \otimes c_2 \otimes c_3 \otimes c_4$
\end{tabular}\]
be a rank 3 tensor.

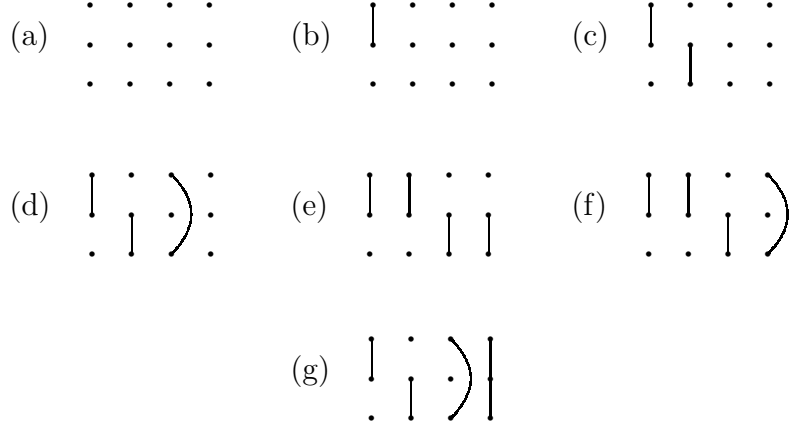
\begin{figure}\caption{Patterns for rank 3 tensors}\label{rankthree}
\begin{center}
\begin{tabular}{lll}
(a) \quad \begin{picture}(65,30)(0,15)
\put(0,0){\circle*{2}}
\put(15,0){\circle*{2}}
\put(30,0){\circle*{2}}
\put(45,0){\circle*{2}}
\put(0,15){\circle*{2}}
\put(15,15){\circle*{2}}
\put(30,15){\circle*{2}}
\put(45,15){\circle*{2}}
\put(0,30){\circle*{2}}
\put(15,30){\circle*{2}}
\put(30,30){\circle*{2}}
\put(45,30){\circle*{2}}
\end{picture} & 
(b) \quad \begin{picture}(65,30)(0,15)
\put(0,0){\circle*{2}}
\put(15,0){\circle*{2}}
\put(30,0){\circle*{2}}
\put(45,0){\circle*{2}}
\put(0,15){\circle*{2}}
\put(15,15){\circle*{2}}
\put(30,15){\circle*{2}}
\put(45,15){\circle*{2}}
\put(0,30){\circle*{2}}
\put(15,30){\circle*{2}}
\put(30,30){\circle*{2}}
\put(45,30){\circle*{2}}
\put(0,15){\line(0,1){15}}
\end{picture} &
(c) \quad \begin{picture}(65,30)(0,15)
\put(0,0){\circle*{2}}
\put(15,0){\circle*{2}}
\put(30,0){\circle*{2}}
\put(45,0){\circle*{2}}
\put(0,15){\circle*{2}}
\put(15,15){\circle*{2}}
\put(30,15){\circle*{2}}
\put(45,15){\circle*{2}}
\put(0,30){\circle*{2}}
\put(15,30){\circle*{2}}
\put(30,30){\circle*{2}}
\put(45,30){\circle*{2}}
\put(0,15){\line(0,1){15}}
\put(15,0){\line(0,1){15}}
\end{picture} \\
(d) \quad \begin{picture}(65,30)(0,15)
\put(0,0){\circle*{2}}
\put(15,0){\circle*{2}}
\put(30,0){\circle*{2}}
\put(45,0){\circle*{2}}
\put(0,15){\circle*{2}}
\put(15,15){\circle*{2}}
\put(30,15){\circle*{2}}
\put(45,15){\circle*{2}}
\put(0,30){\circle*{2}}
\put(15,30){\circle*{2}}
\put(30,30){\circle*{2}}
\put(45,30){\circle*{2}}
\put(0,15){\line(0,1){15}}
\put(15,0){\line(0,1){15}}
\qbezier(30,0)(45,15)(30,30)
\end{picture} &
(e) \quad \begin{picture}(65,60)(0,15)
\put(0,0){\circle*{2}}
\put(15,0){\circle*{2}}
\put(30,0){\circle*{2}}
\put(45,0){\circle*{2}}
\put(0,15){\circle*{2}}
\put(15,15){\circle*{2}}
\put(30,15){\circle*{2}}
\put(45,15){\circle*{2}}
\put(0,30){\circle*{2}}
\put(15,30){\circle*{2}}
\put(30,30){\circle*{2}}
\put(45,30){\circle*{2}}
\put(0,15){\line(0,1){15}}
\put(15,15){\line(0,1){15}}
\put(30,0){\line(0,1){15}}
\put(45,0){\line(0,1){15}}
\end{picture} &
(f) \quad \begin{picture}(65,30)(0,15)
\put(0,0){\circle*{2}}
\put(15,0){\circle*{2}}
\put(30,0){\circle*{2}}
\put(45,0){\circle*{2}}
\put(0,15){\circle*{2}}
\put(15,15){\circle*{2}}
\put(30,15){\circle*{2}}
\put(45,15){\circle*{2}}
\put(0,30){\circle*{2}}
\put(15,30){\circle*{2}}
\put(30,30){\circle*{2}}
\put(45,30){\circle*{2}}
\put(0,15){\line(0,1){15}}
\put(15,15){\line(0,1){15}}
\put(30,0){\line(0,1){15}}
\qbezier(45,0)(60,15)(45,30)
\end{picture} \\\\\\
&(g) \quad \begin{picture}(65,30)(0,15)
\put(0,0){\circle*{2}}
\put(15,0){\circle*{2}}
\put(30,0){\circle*{2}}
\put(45,0){\circle*{2}}
\put(0,15){\circle*{2}}
\put(15,15){\circle*{2}}
\put(30,15){\circle*{2}}
\put(45,15){\circle*{2}}
\put(0,30){\circle*{2}}
\put(15,30){\circle*{2}}
\put(30,30){\circle*{2}}
\put(45,30){\circle*{2}}
\put(0,15){\line(0,1){15}}
\put(15,00){\line(0,1){15}}
\put(45,0){\line(0,1){15}}
\put(45,15){\line(0,1){15}}
\qbezier(30,0)(45,15)(30,30)
\end{picture} & \\\\
\end{tabular}
\end{center}
\end{figure}
We can bring $\psi$ into a reduced form by using $\SL_2$ transformation on each 
qubit. The first three qubits are handled a bit differently than the last one as 
we shall see. We outline how to construct the $g_i \in \SL_2$ that will act on 
each of the first three qubits.

\mbox{\em Case} 1: The set $\{a_i,b_i,c_i\}$ spans $\pH_i$. Assume $\{a_i,c_i\}$ 
is linearly independent and $b_i = \la_i a_i + \mu_i c_i$. With $\la_i \mu_i 
\neq 0$ we can choose $g_i \in \SL_2$ such that $g_i(\la_i a_i) = \nu_i e_0$ and 
$g_i(\mu_i c_i) = \nu_i e_0$ where $\nu_i^2 = \mbox{\rm det} \left[ \la_i a_i | 
\mu_i c_i \right]$. In this situation $g_i(b_i)=\nu_i(e_0+e_1)$. If $\la_i=0$ 
then we can choose $g_i \in \SL_2$ such that $g_i(a_i)=\nu_i e_0$ and $g_i(\mu_i 
c_i)=\nu_i e_1$. A similar argument holds when $\mu_i=0$.

\mbox{\em Case} 2: The set $\{a_i,b_i,c_i\}$ does not span $\pH_i$. We can 
assume that $a_i=b_i=c_i$ and choose $g_i \in \SL_2$ such that $g_i(a_i) = 
g_i(b_i)=g_i(c_i)=e_0$.

Now when $i=4$ the only difference is that if say $\{a_4,b_4\}$ is linearly 
dependent then we cannot simply assume that $a_4=b_4$, but must take into 
account a scalar factor.

Figure \ref{rankthree} contains the different ways that the sets 
$\{a_i,b_i,c_i\}$ can contain the same vector multiple. It is not surprising 
that it is more complicated than Figure \ref{ranktwo} and it is indeed slightly 
more difficult to show that it captures all possibilities. Nevertheless we have:

\begin{proposition}
Any rank $3$ tensor $\psi \in \pH$ is $\SLloc^*$--equivalent to a tensor having 
one of the patterns $(a$$-$$g)$ in Table $\ref{rankthree}$. This pattern is 
uniquely determined by $\psi$.
\end{proposition}
\begin{proof}
Let $\psi=a+b+c$ where \[\begin{tabular}{l}
$a=a_1 \otimes a_2 \otimes a_3 \otimes a_4$, \\
$b=b_1 \otimes b_2 \otimes b_3 \otimes b_4$, \\
$c=c_1 \otimes c_2 \otimes c_3 \otimes c_4$.
\end{tabular}\] If $\psi$ is factorizable, then since it has rank $3$, by the 
classification in \cite{B} it must be $\SLloc^*$--equivalent to a tensor of the 
form $(g)$. Assume $\psi$ is not factorizable. If for each $i$ there are no 
linear dependencies between any two of the factors $a_i$, $b_i$, $c_i$ then 
$\psi$ is clearly in the form $(a)$. If there is one linear dependency, by 
permuting qubits, we may assume $\psi$ is of the form $(b)$. Assume there are 
two linear dependencies. If they are both between the factors of the same two 
summands, say $a$ and $b$, then we may assume that $b=a_1 \otimes a_2 \otimes 
b_3 \otimes b_4$. Now we can use the $\SLloc$--operations as described above to 
get \[\begin{tabular}{lll}
$\nu^{-1}\psi'$ & $=$ & $e_0 \otimes e_0 \otimes e_0 \otimes e_0$ \\
&$+$ & $e_0 \otimes e_0 \otimes (e_0+e_1) \otimes (e_0+e_1)$ \\
&$+$ & $e_1 \otimes e_1 \otimes e_1 \otimes e_1$ \\
& $=$ & $e_0 \otimes e_0 \otimes e_0 \otimes (2e_0+e_1)$ \\
&$+$ & $e_0 \otimes e_0 \otimes e_1 \otimes (e_0+e_1)$ \\
&$+$ & $e_1 \otimes e_1 \otimes e_1 \otimes e_1$ 
\end{tabular}\] which has diagram \[\begin{picture}(65,20)(0,15)
\put(0,0){\circle*{2}}
\put(15,0){\circle*{2}}
\put(30,0){\circle*{2}}
\put(45,0){\circle*{2}}
\put(0,15){\circle*{2}}
\put(15,15){\circle*{2}}
\put(30,15){\circle*{2}}
\put(45,15){\circle*{2}}
\put(0,30){\circle*{2}}
\put(15,30){\circle*{2}}
\put(30,30){\circle*{2}}
\put(45,30){\circle*{2}}
\put(0,15){\line(0,1){15}}
\put(15,15){\line(0,1){15}}
\put(30,0){\line(0,1){15}}
\end{picture}.\] \\ Now we can repeat this process one more time to get a tensor 
in the form $3(e)$. If there is a dependency between a factor of $a$ and $b$, 
and another dependency between a factor of $b$ and $c$, then $\psi$ is in the 
form $(c)$. Assume there are 3 dependencies. If they are between the same two 
summands then $\psi$ is no longer a rank $3$ tensor. If two are between $a$ and 
$b$, and one is between $b$ and $c$, then as before we can bring this into the 
form $(e)$. If we have a dependency between factors of $a$ and $b$, $b$ and $c$, 
and between $a$ and $c$ then $\psi$ is of the form $(d)$. Assume that there are 
4 dependencies. Clearly if at least $3$ are between two summands then $\psi$ is 
not rank $3$. Otherwise it is straightforward to see that the only possibility 
is a tensor of the form $(f)$. 

The uniqueness assertion follows by inspection of Table $\ref{typefamily}$.
\end{proof}

With this in mind, the normal forms follow. 

\begin{remark}\label{irrcomp}
In one of the cases, the proof below depends on the following important fact, a 
special case of \cite[Theorem 38.6.1]{TY}. Let us identify $\pH$ with the 
subspace $\mfp$ of $\gg = \gso_8$. Fix $\psi \in \mfp$ and let $\pO \subset \gg$ 
be its $\SO_8$--orbit under the adjoint action. Then each irreducible component 
of $\pO \cap \mfp$ is a single $\SO_4 \times \SO_4$--orbit. Moreover all these 
components have the same dimension.
\end{remark}

\begin{proposition}\label{rankthreenormal}
A rank $3$ tensor $\psi \in \pH$ of the given pattern (see Figure 
\ref{rankthree}) can be reduced using $\SLloc^*$--operations to the form:\\\\
$(a)$ \quad $\al|0000\rangle +\be(|0\rangle+|1\rangle)^{\otimes 4} +\ga 
|1111\rangle$, $\al\be\ga \neq 0$,  \\
$(b)$ \quad $\al (|0000\rangle + |1111\rangle) + |0\rangle \otimes 
(|0\rangle+|1\rangle)^{\otimes 3}$, $\al \neq 0$, \\
$(c)$ \quad $\al(|0000\rangle + |01\rangle \otimes 
(|0\rangle+|1\rangle)^{\otimes 2} + |1111\rangle)$, $\al \neq 0$, \\
$(d)$ \quad $|0000\rangle + |011\rangle \otimes (|0\rangle+|1\rangle) + 
|1101\rangle$, \\
$(e)$ \quad $\al(|0000 \rangle + |0011 \rangle + |1111 \rangle)$, $\al \neq 0$, 
\\
$(f)$ \quad $|0000\rangle + |0011\rangle + |1110\rangle$, \\
$(g)$ \quad $|0000\rangle + |0110 \rangle + |1100\rangle$.
\end{proposition}

\begin{proof}
Let us outline the procedure in cases $(b)$ and $(d)$. The rest of the cases 
follow from much the same reasoning, although case $(b)$ is uniquely 
non-trivial. Using Figure $\ref{rankthree}$ we see that in the case $(b)$ we may 
assume that \[\psi = \al |0000\rangle + \be |0\rangle \otimes 
(|0\rangle+|1\rangle)^{\otimes 3}+ \ga |1111\rangle \] where $\al\be\ga \neq 0$. 
We apply the $\SL_2$--tranformation $\al e_0 \rightarrow \nu e_0$ and $\ga e_1 
\rightarrow \nu e_1$ to the first qubit to get \[ \al' (|0000\rangle + 
|1111\rangle)+ \be' |0\rangle \otimes (|0\rangle+|1\rangle)^{\otimes 3} \] where 
$\al'=\nu$ and $\be'=\nu\be\al^{-1}$. 

Let \[\psi = \al(|0000\rangle+|1111\rangle) + \be|0\rangle \otimes 
(|0\rangle+|1\rangle)^{\otimes 3}\] with $\al\be \neq 0$ (we rename $\al'$ to 
$\al$ and $\be'$ to $\be$ for convenience). Choose $\ga \in \bC$ such that 
$\ga^2 +\ga=\al(\al+\be)$. If $\be=-\al$ we assume that $\ga=-1$.

We claim that $\psi$ is $\SLloc$--equivalent to \[\phi = \ga(|0000\rangle + 
|1111\rangle)+ |0\rangle \otimes (|0\rangle+|1\rangle)^{\otimes 3}.\] If 
$\be=-\al$ then $\psi = -\al \phi$. Since $\phi$ is nilpotent, our claim holds.

Now assume that $\al+\be \neq 0$. Choose a continuous function $f:[0,1] 
\rightarrow \bC \backslash \{0\}$ such that $f(0)=\al$, $f(1)=\ga$ and $f(t)^2 
\neq \al(\al+\be)$ for all $t \in [0,1]$. Consider the one-parameter tensor 
family \[ \chi (t)=f(t)(|0000\rangle +|1111\rangle) + 
\frac{\al^2-f(t)^2+\al\be}{f(t)}|0\rangle \otimes (|0\rangle+|1\rangle)^{\otimes 
3}. \] It is easy to verify that all matrices $\tilde{R}_{\chi(t)}$ have the 
same Jordan structure: \[0,J_2(\pm i \sqrt{\al^2+\al\be}), J_3(0). \] Hence they 
all belong to a single $\GL_8$--orbit (a similarity class) $\pO \subset M_8$. 
Since $\gg \subset M_8$ is the space of skew-symmetric matrices, $\pO \cap \gg$ 
is a single $\Ort_8$--orbit, and so it is the union of at most two 
$\SO_8$--orbits. By Remark $\ref{irrcomp}$, each irreducible component of $\pO 
\cap \mfp$ is a single $\SO_4 \times \SO_4$--orbit and all these components have 
the same dimension. Since $\{\tilde{R}_{\chi(t)}\}$ is contained in a single 
irreducible component of $\pO \cap \mfp$, and $\chi(0)=\psi$ and $\chi(1)=\phi$, 
we conclude that $\tilde{R}_{\psi}$ and $\tilde{R}_{\phi}$ belong to the same 
$\SO_4 \times \SO_4$--orbit, and so $\psi$ and $\phi$ belong to the same 
$\SLloc$--orbit. This concludes the proof of our claim.

A tensor $\psi$ of the form $(d)$ can be reduced to \[ \al |0000\rangle + \be 
|011\rangle \otimes (|0\rangle+|1\rangle) + \ga|1101\rangle. \] By mapping $\al 
e_0 \rightarrow \nu e_0$ and $\ga e_1 \rightarrow \nu e_1$ in the third qubit we 
attain \[ \al' |0000\rangle + \be' |011\rangle \otimes (|0\rangle+|1\rangle) + 
\al'|1101\rangle \] and we can check that $\al'=\nu$ and $\be'=\nu\be\al^{-1}$. 
Now by applying the $\SL_2$--tranformation which sends $\al' e_0 \rightarrow 
\nu' e_0$ and $\be' e_1 \rightarrow \nu' e_1$ in the third qubit we obtain \[ 
\nu'(|0000\rangle + |011\rangle \otimes (|0\rangle+|1\rangle) + |1101\rangle). 
\] Since $\psi$ is nilpotent, the $\SLloc^*$ and $\GLloc^*$--orbits coincide and 
we can replace $\nu'$ with $1$.
\end{proof}

We shall say that the tensors listed in Proposition \ref{rankthreenormal} are of 
type 3$(a$$-$$g)$, respectively. Note that the invariants $L$ and $M$ vanish on 
each of these tensors.

For the computation of tensor ranks it is important to know the Jordan structure 
of the matrices $\tilde{R}_\psi$ for all types of tensors of ranks $\leq 3$. We 
shall investigate the tensor $\psi$ of type 3 $(a)$ in detail. The other cases 
are easy to analyze and we omit their discussion. By using (\ref{charpoly}) we 
find that the characteristic polynomial of $\tilde{R}_{\psi}$ is 
\[t^2(t^6+2(\al\be+\al\ga+\be\ga)t^4+(\al\be+\al\ga+\be\ga)^2t^2+4(\al\be\ga)^2)
. \] If we let $s=t^2$ then it is $sg(s)$ where 
\[g(s)=s^3+2(\al\be+\al\ga+\be\ga)s^2+(\al\be+\al\ga+\be\ga)^2s+4(\al\be\ga)^2.
\]
The discriminant of $g(s)$ (see (\ref{discr})) is 
\[16(\al\be\ga)^2((\al\be+\al\ga+\be\ga)^3-27(\al\be\ga)^2).\] If the 
discriminant does not vanish then $\tilde{R}_{\psi}$ is semisimple and belongs 
in family $1$ from Table $\ref{orbitreps}$. Now if it vanishes then we must have 
\begin{equation}\label{disc}(\al\be+\al\ga+\be\ga)^3-27(\al\be\ga)^2=0
\end{equation} 
and the roots of $g(s)$ are 
\begin{equation}\label{roots}\la^2=-\frac{4}{3}(\al\be+\al\ga+\be\ga), \quad 
\mu^2=-\frac{1}{3}(\al\be+\al\ga+\be\ga), \end{equation} where the latter is a 
double root.

Now we can determine the Jordan structure of $\tilde{R}_{\psi}$. Since the 
eigenvalue $0$ of $\tilde{R}_{\psi}$ has multiplicity two, and the matrix rank 
of $\tilde{R}_{\psi}$ is 6 (there are two rows of zeros) we conclude that there 
are two $1 \times 1$ Jordan blocks of $0$. Consider the matrix 
$\tilde{R}_{\psi}^2$. By permuting rows and columns, we see that it is similar 
to a matrix of the form $P_1 \oplus 0 \oplus P_2$ where the second summand is a 
$2 \times 2$ zero matrix and the other two summands are $3 \times 3$ matrices. 
It is straightforward to see that $P_1$ and $P_2$ are similar. Now assume that 
$\tilde{R}_{\psi}$ is semisimple, then so is $\tilde{R}_{\psi}^2$. In particular 
the matrices $P_1-\la^2I_3$ and $P_1-\mu^2I_3$ have ranks 1 and 2 respectively. 
But then by evaluating the $2 \times 2$ minors we can conclude that $\al = \be = 
\ga$. Conversely one can check that $\al = \be = \ga$ implies that 
$\tilde{R}_{\psi}$ is semisimple. We find that $\tilde{R}_\psi$ then has the 
Jordan structure: $0$,$0$,$\pm \la$, $J_2(\pm \mu)$ and it is in family $2$, 
unless $\al=\be=\ga$ in which case $\tilde{R}_\psi$ is semisimple and is again 
in family $1$.

To summarize, we find that if a tensor of rank 3 is semisimple, then it must be 
of type $3(a)$. Furthermore, a rank 3 tensor of type $(a)$ is semisimple unless 
(\ref{disc}) holds and $\al$, $\be$, $\ga$ are not all equal.

Table \ref{typefamily} presents the Jordan structures for the different type of 
tensors of rank $\leq 3$. Note that one has to permute the four qubits in order 
to obtain all possible Jordan structures for a given type.

\begin{table}[h]\caption{Jordan structure of $\tilde{R}_{\psi}$ for tensors of 
rank $\leq 3$}\label{typefamily}
\begin{center}
\begin{tabular}{ll}
$1(a)$ & $0$,$0$,$0$,$0$,$J_2(0)$,$J_2(0)$. \\\\
$2(a)$ & $0$,$0$,$0$,$0$,$\pm i\sqrt{\al}$,$\pm i\sqrt{\al}$. \\
$2(b)$ & $0$,$0$,$J_3(0)$,$J_3(0)$. \\
$2(c)$ & $0$,$0$,$0$,$0$,$0$,$J_3(0)$ and $J_2(0)$,$J_2(0)$,$J_2(0)$,$J_2(0)$. 
\\\\
$3(a)$ & Discussed above: Semisimple or $0$,$0$,$\pm i\la$,$J_2(\pm i\mu)$. \\
$3(b)$ & If $\al=-1$ then $0,J_7(0)$ else $0, J_2(\pm i\sqrt{\al^2 + \al}), 
J_3(0)$. \\
$3(c)$ & $\pm i\al, \pm i\al,0,J_3(0)$ and $J_2(0)$,$J_2(0)$,$J_2(\pm i\al)$. \\
$3(d)$ & $J_3(0),J_5(0)$. \\
$3(e)$ & $\pm i\al, \pm i\al,J_2(0),J_2(0)$ and $0$,$0$,$0$,$0$,$J_2(\pm i\al)$. 
\\
$3(f)$ & $0,0,0,J_5(0)$ and $J_4(0),J_4(0)$. \\
$3(g)$ & $0,J_2(0),J_2(0),J_3(0)$. \\
\end{tabular}
\end{center}
\end{table}

\section{Determining the tensor ranks}\label{tensorranks}
Here we will compute the ranks of tensors $\psi$ in each of the nine families 
listed in Theorem \ref{classify II}. For any subsequence $\{a_i\}$ of $1,2,3,4$ 
we set $\pH_{a_1 \cdot\cdot\cdot a_k}=\bigotimes_{i=1}^k \pH_{a_i}$. For 
$\psi=e_0 \otimes t_0 + e_1 \otimes t_1$, where $\{t_0,t_1\}$ is linearly 
independent, we form the linear transformation $T_{\psi}:\bC^2 \rightarrow 
\pH_{234}$ sending $(x,y) \rightarrow x t_0 + y t_1$. The image of $\bC^2 
\backslash \{0\}$ under $T_{\psi}$ is a projective line $l_{\psi}$ in the 
projective space $\bP(\pH_{234})$. For a pure state $\phi \in \pH_{234}$ we will 
denote by $\det \phi$ the Cayley hyperdeterminant as described in \cite{B}. 
Explicitly we have \[ \det \phi = (\tr{A}\, \tr{B}-\tr{AB})^2-4 \det{A} \det{B} 
\] where \[A=\begin{bmatrix} \phi_{000} & \phi_{001} \\ \phi_{010} & \phi_{011} 
\end{bmatrix}; \quad B= \begin{bmatrix} \phi_{100} & \phi_{101} \\ \phi_{110} & 
\phi_{111} \end{bmatrix} .\] Note that $4 |\det \phi|$ is the residual 
entanglement (also known as the $3$--tangle) of the pure state $\phi$ described 
in \cite{CKW}.

\begin{lemma} \label{lem}
If $(x_1,y_1)$ and $(x_2,y_2)$ are linearly independent then we have that 
$\rank\psi\leq \rank T_{\psi}(x_1,y_1)+\rank T_{\psi}(x_2,y_2)$.
\end{lemma}

\begin{proof}
Since $\{(x_1,y_1),(x_2,y_2)\}$ is linearly independent, so is $\{x_1 e_0 + x_2 
e_1,y_1 e_0 + y_2 e_1\}$.  The tensor $\psi '=(x_1 e_0 + x_2 e_1) \otimes t_0 + 
(y_1 e_0 + y_2 e_1) \otimes t_1$ is in the same $\GLloc$--orbit as $\psi$ so 
$\rank\psi '=\rank\psi$.  But $\psi '=e_0 \otimes (x_1 t_0 + y_1 t_1) + e_1 
\otimes (x_2 t_0 + y_2 t_1)$ so $\rank\psi '\leq \rank T_{\psi}(x_1,y_1)+\rank 
T_{\psi}(x_2,y_2)$ and the result follows.
\end{proof}

\begin{table}[t]\caption{The $\SLloc$--invariants 
$H,L,M,N,D,\Gamma$}\label{polyinv}
\begin{enumerate}[]
\item \begin{enumerate}[1.]
\item $\frac{1}{2}(a^2+b^2+c^2+d^2)$ ; $abcd$ ; 
$\frac{1}{16}(4(ad-bc)^2-(a^2-b^2-c^2+d^2)^2)$; \\ 
$-\frac{1}{16}(4(ac+bd)^2-(a^2-b^2+c^2-d^2)^2)$ ; 
$-\frac{1}{4}(ad-bc)(ab-cd)(ac-bd)$; \\ 
$\frac{1}{32}(2(a^6+b^6+c^6+d^6)-(a^2+b^2+c^2+d^2)(a^4+b^4+c^4+d^4)+18(a^2b^2c^2
+b^2c^2d^2+c^2d^2a^2+d^2a^2b^2))$
\item $\frac{1}{2}(a^2 +b^2+2c^2)$ ; $abc^2$ ; 
$-\frac{1}{16}(a-b)^2((a+b)^2-4c^2)$; $\frac{1}{16}(a+b)^2((a-b)^2-4c^2)$ ; \\  
$\frac{1}{4}c^2(a-b)^2(c^2-ab)$; 
$\frac{1}{32}((a^2+b^2)((a^2-b^2)^2+16c^4)-2c^2(a^4-18a^2b^2+b^4))$
\item $\frac{1}{2}(a^2+b^2)$ ; $0$ ; $-\frac{1}{16}(a^2-b^2)^2$ ; 
$\frac{1}{16}(a^2-b^2)^2$ ; $0$ ; $\frac{1}{32}(a^2-b^2)(a^4-b^4)$
\item $\frac{1}{2}(a^2+b^2)$ ; $0$ ; $-\frac{1}{16}(a^2-b^2)^2$ ; 
$\frac{1}{16}(a^2-b^2)^2$ ; $0$ ; $\frac{1}{32}(a^2-b^2)(a^4-b^4)$
\item $a^2+b^2$ ; $a^2b^2$ ; $0$ ; $-a^2b^2$ ; $0$ ; $a^2b^2(a^2+b^2)$
\item $\frac{1}{2}(a^2+3b^2)$ ; $ab^3$ ; $\frac{1}{16}(b-a)^3(a+3b)$; 
$\frac{1}{16}(a-3b)(a+b)^3$ ; $\frac{1}{4}b^3(b-a)^3$; \\ 
$\frac{1}{32}((a^2-b^2)^3+16b^4(3a^2+b^2))$
\item $\frac{1}{2}a^2$ ; $0$ ; $-\frac{1}{16}a^4$ ; $\frac{1}{16}a^4$ ; $0$ ; 
$\frac{1}{32}a^6$
\item $\frac{1}{2}a^2$ ; $0$ ; $-\frac{1}{16}a^4$ ; $\frac{1}{16}a^4$ ; $0$ ; 
$\frac{1}{32}a^6$
\item $2a^2$ ; $a^4$ ; $0$ ; $-a^4$ ; $0$ ; $2a^6$
\item $a^2$ ; $0$ ; $0$ ; $0$ ; $0$ ; $0$
\item $a^2$ ; $0$ ; $0$ ; $0$ ; $0$ ; $0$
\end{enumerate}
\item $^*$ Families 12 - 17 are nilpotent so all invariants are 0.
\end{enumerate}
\end{table}

Table \ref{polyinv} lists, in order, the invariants $H$,$L$,$M$,$N$, $D$ and 
$\Gamma$ for each of the $17$ families from Table \ref{orbitreps}. It will be 
helpful to refer to the list of invariants of the families as we proceed. We 
will often use the fact that if either $L$ or $M$ does not vanish on $\psi$ then 
$\psi \not \in \bar{S_3}$ and so $\psi$ must have rank $4$ (see Proposition 
\ref{rankfour}). It is easy to determine whether a state $\psi$ is factorizable. 
For instance, we have $\psi = \phi \otimes \chi$ with $\phi \in \pH_{12}$ and 
$\chi \in \pH_{34}$ iff $\rank \tilde{\psi}=1$. Similarly we have $\psi = \phi 
\otimes \chi$ with $\phi \in \pH_1$ and $\chi \in \pH_{234}$ iff the $2 \times 
8$ matrix $[\psi_{i,jkl}]$ has rank $1$. In general, one has first to permute 
the qubits.

We shall now consider separately each of the nine families mentioned in Theorem 
$\ref{classify II}$.\\
\\
\mbox{\em Family 1}. We may permute the diagonal entries $a$,$b$,$c$,$d$ of $R$ 
(see Table $\ref{orbitreps}$) and replace them by $\pm a$,$\pm b$,$\pm c$,$\pm 
d$ without changing the $\SLloc^*$--orbit of $\psi$.
If $L \neq 0$ or $M \neq 0$ then $\rank \psi =4$. From now on we may assume that 
$a+b+c=d=0$ (see the expressions for $L$ and $M$ in Table $\ref{polyinv}$). Then 
we have \[\begin{tabular}{lll}
$4\psi$ & $=$ & $a(e_0+e_1) \otimes (e_0+e_1) \otimes (e_0-e_1) \otimes 
(e_0-e_1)$ \\
&$+$ & $a(e_0-e_1) \otimes (e_0-e_1) \otimes (e_0+e_1) \otimes (e_0+e_1)$ \\
&$-$ & $4c(e_0 \otimes e_1 \otimes e_1 \otimes e_0 + e_1 \otimes e_0 \otimes e_0 
\otimes e_1)$.
\end{tabular}\]
If $abc=0$, say $c=0$, then $\rank \psi =2$. Note that after normalization, this 
$\psi$ represents the generalized $GHZ$ state. Now we may assume that $abc \neq 
0$. Let $\phi$ be the tensor of type $3(a)$ (see Proposition 
\ref{rankthreenormal}). We shall choose the scalars $\al, \be, \ga$ to satisfy 
the two equations
\begin{equation}\label{charinv} 2(\al \be + \be \ga + \ga \al) = a^2 + b^2+ c^2, 
\quad 4(\al \be \ga)^2 = a^2 b^2c^2.\end{equation} and to ensure that $\phi$ is 
semisimple. If $a$, $b$, $c$ are not distinct, say $a=b$, we can take 
$\al=\be=\ga=a$. For semisimplicity of $\phi$ in this case see the end of the 
previous section.

From now on we assume that $a$,$b$, and $c$ are distinct. If $a^2+b^2+c^2=0$ we 
take $\be=\al\zeta$ and $\ga = \al \zeta^2$; where $\zeta = e^{2\pi i /3}$, and 
we choose $\al$ such that $-2\al^3 =abc$. Otherwise we take $\be=-\al$, 
$\ga=\frac{abc}{2\al^2}$ and choose $\al$ such that $-2\al^2=a^2+b^2+c^2$. The 
equations ($\ref{charinv}$) imply that $\tilde{R}_{\psi}$ and $\tilde{R}_{\phi}$ 
have the same characteristic polynomial. Since the nonzero eigenvalues of 
$\tilde{R}_{\psi}$, i.e. $\pm ia$,$\pm ib$,$\pm ic$, are distinct, one can 
verify easily that $\tilde{R}_{\phi}$ is semisimple. As $L$ and $M$ vanish on 
$\psi$ and $\phi$ and the equations ($\ref{charinv}$) show that $H$ and $D$ also 
agree on $\psi$ and $\phi$, Theorem $\ref{orbit}$ implies that $\psi$ and $\phi$ 
are in the same $\SLloc$--orbit. Hence $\rank \psi = \rank \phi = 3$. \\
\\
\mbox{\em Family 2}. If $\psi \not \in \bar{S_3}$ then $\rank\psi=4$. Otherwise 
we must have $L=M=0$, i.e., $abc=(a-b)((a+b)^2-4c^2)=0$. As we may switch $a$ 
and $b$ and multiply $a,b,c$ by $\pm 1$, there are only three cases to consider:
\begin{enumerate}[(i)]
\item $a=b$, $c=0$;
\item $a=b=0$, $c\neq 0$;
\item $b=0$, $a=-2c\neq 0$.
\end{enumerate}
In case $(\rm i)$ we have \[\begin{tabular}{lll}
$4\psi$ & $=$ & $(e_0-e_1) \otimes (e_0+e_1) \otimes (e_0+e_1) \otimes 
((a-2i)e_0+(a+2i)e_1)$ \\
&$+$ & $a(e_0-e_1) \otimes (e_0-e_1) \otimes (e_0-e_1) \otimes (e_0-e_1)$ \\
&$+$ & $2ae_1 \otimes (e_0+e_1) \otimes (e_0+e_1) \otimes (e_0+e_1)$, \\
\end{tabular}\] and in case $(\rm ii)$ \[\begin{tabular}{lll}
$4\psi$ & $=$ & $c(e_0+e_1) \otimes (e_0-e_1) \otimes (e_0+e_1) \otimes 
(e_0-e_1)$ \\
&$+$ & $c(e_0-e_1) \otimes (e_0+e_1) \otimes (e_0-e_1) \otimes (e_0+e_1)$ \\
&$-$ & $2i(e_0-e_1) \otimes (e_0+e_1) \otimes (e_0+e_1) \otimes (e_0-e_1)$. \\
\end{tabular}\] Clearly if $a=0$ in case $(\rm i)$ then $\rank\psi=1$. Otherwise 
in cases $(\rm i)$ and $(\rm ii)$ it is easy to verify that $\psi \not \in 
\bar{S_2}$ and so $\rank\psi=3$. Now we consider case $(\rm iii)$. Let $\phi \in 
\pH$ be of type 3$(a)$ with $\al = \frac{c}{3}$, $\be = ic\sqrt{3}$ and $\ga 
=-\be$. Then the matrices $R_{\psi}$ and $PR_{\phi}$, where $P$ is the diagonal 
matrix with $(-1,1,1,1)$ diagonal entries, are symmetric and have the same 
Jordan structure: $0$, $-2c$, $J_2(c)$. Hence they are orthogonally similar. 
This shows that $R_{\psi}$ and $R_{\phi}$ are in the same $\Ort_4 \times 
\Ort_4$--orbit. Hence $\psi$ and $\phi$ are in the same $\SLloc^*$--orbit and so 
$\rank \psi = \rank \phi = 3$.\\
\\
\mbox{\em Family 3}. If $a^2 \neq b^2$ then $\psi \not \in \bar{S_{3}}$ and 
$\rank\psi=4$. Since we can interchange $a$ and $b$ and replace them by $\pm a$ 
and $\pm b$, we may assume that $a=b$. If $a=0$ then \[2\psi=(e_0+e_1) \otimes 
(e_0-e_1) \otimes (e_1 \otimes e_0 - e_0 \otimes e_1) \] and $\rank \psi =2$. If 
$a \neq 0$ then $\psi \not \in \bar{S_2}$ and \[\begin{tabular}{lll}
$4a\psi$ & $=$ & $(e_0+e_1) \otimes ((a^2+1)e_0+(a^2-1)e_1) \otimes (e_0+e_1) 
\otimes (e_0+e_1)$ \\
&$-$ & $(e_0+e_1) \otimes (e_0-e_1) \otimes ((1+a)e_0+(1-a)e_1)$ \\ 
&&\quad \quad \quad \quad $\otimes ((1-a)e_0+(1+a)e_1)$ \\
&$-$ & $2a^2e_1 \otimes (e_0-e_1) \otimes (e_0-e_1) \otimes (e_0-e_1)$, \\
\end{tabular}\] and so $\rank \psi =3$. \\
\\
\mbox{\em Family 6}. If $\psi \not \in \bar{S_3}$ then $\rank \psi = 4$. 
Otherwise $L=M=0$ and by using Table \ref{polyinv}, we have $a=b=0$. Then $\psi$ 
is nilpotent and it is easy to verify that $\psi \in \bar{S_2}$. Since $\psi$ is 
not factorizable, it cannot have $\rank$ $1$ or $2$ (see Proposition 
\ref{ranktwonormal}). The matrix $\tilde{R}_{\psi}$ has Jordan structure $0$, 
$0$, $J_3(0)$, $J_3(0)$ (see Table \ref{orbitreps}). Since this is absent from 
the rank $3$ section of Table $\ref{typefamily}$, we infer that $\rank \psi \neq 
3$. Thus $\rank \psi =4 $. \\
\\
\mbox{\em Family 9}. If $a \neq 0$ then $\psi \not \in \bar{S_{3}}$ and so 
$\rank\psi=4$. If $a=0$ then \[ \psi = 2i(|1110\rangle - |0010\rangle + 
|1001\rangle) \] and $\psi \not \in \bar{S_2}$ so $\rank\psi=3$. \\
\\
\mbox{\em Family 10}. Permute qubits 1 and 2. (The effect on the matrix $R_\psi$ 
is just to change its $(3,3)$--entry from $1$ to $-1$.) If $a \neq 0$ then 
$\rank T(1,1)=1$ and $\rank T(1,0)=2$.  Since $\psi \not \in \bar{S_{2}}$ we 
have that $\rank\psi=3$. If $a=0$ then $\psi=(e_1-e_0) \otimes \phi$. Since 
$\det \phi = 0$ and $\phi$ is not factorizable, we have $\rank \psi = \rank \phi 
=3$. \\
\\
\mbox{\em Family 12}. We check that $\psi \not \in \bar{S_2}$ since the 1,1 
minor of $\tilde{\psi}$ is nonzero. We have that $\rank T(1,0)=1$ and $\rank 
T(1,1)=2$. Hence $\rank\psi=3$. A computation gives \[\begin{tabular}{lll}
$(1-i)\sqrt{2}\psi$ & $=$ & $(1-i)\sqrt{2}(e_0-e_1) \otimes e_1 \otimes 
(e_0-ie_1) \otimes (-ie_0+e_1)$ \\
&$-$ & $e_1 \otimes (e_0-i\sqrt{2}e_1) \otimes (e_0+\be e_1) \otimes (e_0+\al 
e_1)$ \\
&$-$ & $e_1 \otimes (e_0+i\sqrt{2}e_1) \otimes (-e_0+\al e_1) \otimes (e_0-\be 
e_1)$, \\
\end{tabular}\] where $\al= \sqrt{2}+1$ and $\be = \sqrt{2}-1$. \\
\\
\mbox{\em Family 14}. It is again straightforward to verify that $\psi \not \in 
\bar{S_2}$. Then using that $\rank T(i,1)=1$ and $\rank T(1,0)=2$, we obtain 
that $\rank \psi =3$. \\
\\
\mbox{\em Family 16}. In this case $\psi = (e_0+e_1) \otimes \phi$. Since $\det 
\phi \neq 0$ we have $\rank \psi = \rank \phi = 2$. \\

After all these computations it is worthwhile observing that $\bar{S_3}$ 
contains only one $\SLloc^*$--orbit of rank 4 tensors. This exceptional orbit is 
the unique nilpotent orbit of family $6$ (with $a=b=0$). It is the orbit of the 
generalized $W$ state given by (\ref{Wstate}) and it is contained in 
$\bar{S_2}$. In particular we have $\bar{S_2} \neq S_2$ and $\bar{S_3} \neq 
S_3$.

\section{Tensor rank algorithm}\label{algo}

By using the results of the previous section, we can now construct a simple 
algorithm for computing the tensor rank. We have explained in the previous 
section how to test a state $\psi$ for factorization. If one of the factors is 
from a single $\pH_k$, we may use density matrices. For a state $\psi \in \pH$ 
let $\rho = | \psi \rangle \langle \psi |$ be its density matrix. Denote by 
$\rho_k$ its reduced density matrix obtained by tracing out all qubits but the 
$k$-th one (for the definition of the density matrices and partial
trace see e.g. \cite{JP}). 
Then $\psi$ factorizes, with one of the factors in 
$\pH_k$, iff the matrix $\rho_k$ has rank $1$, so we let $r_k$ be the matrix 
rank of $\rho_k$. With an abuse of notation, we let $r_k$ be the rank of the 
corresponding $\rho_k$ for $3$-qubit tensors as well.

We now give our algorithm for computing the tensor rank of an arbitrary state 
$\psi \in \pH$. The algorithm uses another procedure which computes the tensor 
rank of $3$--qubit states, which can be deduced from \cite{B}. It should be 
understood that the algorithms halt as soon as the rank is returned. Recall the 
definition of the hyperdeterminant $\det \psi$ for $3$--qubit pure states $\psi$ 
given in the previous section.

\subsection*{3--Qubit Tensor Rank Algorithm}

\begin{enumerate}[]
\item \mbox{\tt \,\,\,Input: } \mbox{A nonzero tensor $\psi \in \pH_{123}$}
\item \mbox{\tt Output: } \mbox{The tensor rank of $\psi$}\\
\end{enumerate}

\begin{enumerate}[1]
\tt{
\item If $\det \psi$ is nonzero then return $2$.
\item Compute the ranks $r_k$ of $\rho_k$ for $k \in \{1,2,3\}$.
\item If $r_k=1$ for at least two different $k$ then return $1$.
\item If some $r_k=1$ then return $2$.
\item Return $3$.
}
\end{enumerate}

With this we may compute ranks of $4$--qubit tensors.

\subsection*{$4$-Qubit Tensor Rank Algorithm}

\begin{enumerate}[]
\item \mbox{\tt \,\,\,Input: } \mbox{A nonzero tensor $\psi \in \pH$}
\item \mbox{\tt Output: } \mbox{The tensor rank of $\psi$}\\
\end{enumerate}

\begin{enumerate}[1]
\tt{
\item If $L(\psi)$ or $M(\psi)$ is nonzero then return $4$.
\item If at least one of the forty-eight $3 \times 3$ minors of the \\ matrices 
$\tilde{\psi}$, $\tilde{\psi}'$, $\tilde{\psi}''$ is nonzero then return $3$.
\item Compute the ranks $r_k$ of $\rho_k$.
\item If say $r_1 =1$, then $\psi =v_1 \otimes \phi$ with $v_1 \in \pH_1$ and 
$\phi \in \pH_{234}$, and return $\rank \phi$.
\item Now all $r_k=2$. If $\psi$ is nilpotent, i.e., $\tilde{R}_\psi$ is 
nilpotent, then return $4$.
\item Return $2$.
}
\end{enumerate}

Let us show that the algorithm is correct. It may be helpful to look at Figure 
\ref{rankpic} where some of the sets we use below are exhibited. Step $1$ is 
clear. In order to justify step $2$, it suffices to verify that \[ \psi \in 
\bar{S_3} \backslash \bar{S_2} \Rightarrow \rank \psi =3. \] This follows from 
the case-by-case analysis of the previous section.

After reaching Step $3$, we have $\psi \in \bar{S_2}$. Consequently, the 
families $7$,$8$,$9$,$12$,$13$,$14$ and $15$ are ruled out, i.e., $\psi$ does 
not belong to any of them. Indeed it is easy to verify that none of these 
families meets $\bar{S_2}$.

Steps $3$ and $4$ are also clear.

After reaching step $5$ our $\psi$ is non-factorizable and we can rule out the 
families $2$,$3$,$4$,$5$,$10$,$11$,$16$ and $17$. Indeed the families $16$ and 
$17$ are factorizable and the orbits in the families $2$,$3$,$4$,$5$,$10$ and 
$11$ which are contained in $\bar{S_2}$ are also factorizable. Hence $\psi$ 
belongs to the family $1$ or $6$. If it is nilpotent, it is in family $6$ and 
has rank $4$. Otherwise it is in family $1$ and the detailed analysis of this 
case in the previous section shows that the rank of $\psi$ is $2$.

Figure \ref{rankpic} describes the structure of $\pH$ with respect to tensor 
ranks. Each vertex represents a Zariski closed set and it is ordered by 
inclusion as one progresses to the top vertex. All sets on or below the 
horizontal line satisfy the equation $H=0$ and consist of nilpotent orbits. 
However there exist nilpotent orbits not contained in $\bar{S_2}$. The numbers 
between pairs of adjacent vertices indicate the rank of the tensors that are in 
the set theoretic difference between the higher and lower vertices. The numbers 
on the far left indicate the dimension of the corresponding affine variety.

\begin{figure}[hbtp]\caption{Some subvarieties of $\pH$}\label{rankpic}
\begin{center}
\begin{picture}(200,370)(0,0)
\put(100,350){\circle*{2}}
\put(100,300){\circle*{2}}
\put(100,250){\circle*{2}}
\put(50,200){\circle*{2}}
\put(150,200){\circle*{2}}
\put(100,150){\circle*{2}}
\put(100,100){\circle*{2}}
\put(100,50){\circle*{2}}
\put(100,0){\circle*{2}}

\put(100,350){\line(0,-1){50}}
\put(100,300){\line(0,-1){50}}
\put(100,250){\line(1,-1){50}}
\put(100,250){\line(-1,-1){50}}
\put(50,200){\line(1,-1){50}}
\put(150,200){\line(-1,-1){50}}
\put(25,200){\line(1,0){150}}
\put(100,150){\line(0,-1){50}}
\put(100,100){\line(0,-1){50}}
\put(100,50){\line(0,-1){50}}

\put(80,350){$\pH$}

\put(0,350){$16$}
\put(0,300){$14$}
\put(0,250){$10$}
\put(0,200){$9$}
\put(0,150){$8$}
\put(0,100){$6$}
\put(0,50){$5$}
\put(0,0){$0$}

\put(80,325){\framebox{$4$}}
\put(80,275){\framebox{$3$}}
\put(95,230){\framebox{$2$}}
\put(50,170){\framebox{$4$}}
\put(140,170){\framebox{$2$}}
\put(80,120){\framebox{$3$}}
\put(80,70){\framebox{$2$}}
\put(80,20){\framebox{$1$}}

\put(80,300){$\bar{S_3}$}\put(110,300){$L=M=0$}
\put(80,250){$\bar{S_2}$}\put(110,250){$48$ equations}
\put(35,200){$\bar{W}$}\put(180,200){$H=0$}
\put(80,50){$\bar{S_1}$}

\put(110,150){$3 \cdot 2^2 \cdot 1$}
\put(110,100){$2^4; 3 \cdot 1^5$}
\put(110,50){$2^2 \cdot 1^4$}

\put(110,0){\bf 0}
\end{picture}
\end{center}
\end{figure}

\section{Conclusion}\label{conc}

In this paper we have investigated the SLOCC classification of pure states of 
four qubits first described by Verstraete et al. in \cite{VDDV}. The families of 
representatives provided in that paper were accurate except for some possible 
misprints in the family $L_{ab_3}$. However their claim of uniqueness in Theorem 
1 is not true and the subsequent proof was not easy to follow.

We have provided a more general version of that theorem in our Theorem 
\ref{indecompreps}. We presented this theorem within the framework of orthogonal 
representations of a certain quiver $\pQ$. We felt that this approach lead to a 
simpler proof of Theorem \ref{indecompreps}.  We also observed that this theorem 
can be deduced (with some additional work) from the theory of symmetric quivers 
as presented in a recent paper of Derksen and Weyman \cite{DW}.

We found it beneficial to embed the $4$--qubit Hilbert space $\pH$ into the Lie 
algebra $\gg$ of the complex orthogonal group $\O_8$. This naturally lead to the 
notion of semisimple and nilpotent states. The semisimple states are dense in 
$\pH$ while the nilpotent ones comprise only finitely many $\SLloc$--orbits. The 
subgroup $\Ort_4 \times \Ort_4 \subset \Ort_8$ acts naturally on $\pH$.

We have also provided a more complete description of the behaviour of the 
$\Ort_4 \times \Ort_4$--orbits under permutations of qubits. This was addressed 
in \cite{VDDV} as well, but an important characteristic was not stressed. Namely 
that the action of a permutation of qubits on a family of orbits may map some 
orbits into a different family while at the same time map others back into 
itself. So in general, a permutation of qubits does not induce a permutation of 
families of orbits as naivete would have one think.

The problem of showing that two states $\phi,\psi \in \pH$ are not 
$\SLloc^*$--equivalent appears in a number of recent papers \cite{BR,WYKO}. We 
derived polynomial invariants for $\SLloc^*$ and show that two semisimple states 
are $\SLloc^*$--equivalent iff they agree on the invariants. In 1852, these same 
invariants were considered by Schl\"{a}fli who noted their invariance under 
permutations of indices. The general case is somewhat more complicated as it 
requires computing the Jordan structure of associated matrices $\tilde{R}$, and 
possibly the use of $ab$--diagrams.

The other focus of this paper was to ultimately develop an algorithm that would 
calculate the tensor rank for a nonzero tensor in $\pH$. This was accomplished 
by a thorough examination of each of the families in the $\SLloc^*$ 
classification. To carry out this analysis we used the results of Brylinski in 
\cite{B}. In particular it was essential to know that the maximum rank of a 
tensor in $\pH$ is $4$ and that the polynomial $\SLloc$ invariants $L$ and $M$ 
define the Zariski closure of the tensors of rank $\leq 3$. We found that 
another set of 48 equations define the Zariski closure of the tensors of rank 
$\leq 2$ based on the speculation by Brylinski in \cite{B}. It was also 
fortunate that the tensors of rank $\leq 3$ admitted a simple classification 
which allowed us to deduce the ranks of some other tensors in certain cases. We 
then were able to construct the algorithm which is pleasantly simple compared to 
the machinery involved in the analysis mentioned above.

The authors of \cite{VDDV} claim to have solved the problem of equivalence of 
two states under the group $\rm U_{loc}$ of local unitary operations. For that 
purpose they propose a normal form via a two step procedure. Apparently they 
failed to observe that the second step of their procedure may undo the 
beneficial effect of the first step. It is unclear how their proposed normal 
form is actually defined. Since $\rm U_{loc}$ is a compact group, this 
equivalence problem can be solved by computing the algebra of real polynomial 
invariants $\pH \rightarrow \bR$ for $\rm U_{loc}$. Indeed it is known in 
general that these invariants separate $\rm U_{loc}$--orbits. However, a set of 
generators for this algebra of invariants has not been computed so far although 
its Poincar\'{e} series has been computed independently in \cite{W} and 
\cite{LTT}. The problem of local unitary equivalence is also considered in 
\cite{AM} but the results are far from conclusive. Hence this problem remains 
open.

\appendix

\section*{Appendix A}

The following table gives the representatives $\psi$ of the nine families of 
$\SLloc^*$--orbits of non-normalized pure states of four qubits. They are 
derived from the corresponding $R$--matrices given in Table \ref{orbitreps} by 
the transformation $R_\psi \rightarrow \tilde{\psi} = T^{-1} R_\psi T$. We point 
out that our $R$--matrices for these families are chosen to be as simple as 
possible, consequently the expressions for the corresponding states $\psi$ are 
not. The representations given in \cite{VDDV} are in some cases shorter than our 
representations, e.g. for family 16 their representative is \[ |0\rangle \otimes 
(|000\rangle+|111\rangle). \]

\begin{table}[htbp]\caption{Representatives $\psi$ of 
$\SLloc^*$--orbits}\label{psiexpr}
\begin{tabular}{ll}
\\
1. & $\frac{a+d}{2}(|0000\rangle + |1111\rangle) + \frac{a-d}{2}(|0011\rangle + 
|1100\rangle)$ \\
& $+\frac{b+c}{2}(|0101\rangle+|1010\rangle)
+\frac{b-c}{2}(|0110\rangle + |1001\rangle)$ \\\\
2. &  
$\frac{a+c-i}{2}(|0000\rangle+|1111\rangle)+\frac{a-c+i}{2}(|0011\rangle+|1100
\rangle)$ \\ 
&$+\frac{b+c+i}{2}(|0101\rangle+|1010\rangle)
+\frac{b-c-i}{2}(|0110\rangle+|1001\rangle)$ \\ & $
+\frac{i}{2}(|0001\rangle+|0111\rangle+|1000\rangle+|1110\rangle $ \\ & $
-|0010\rangle-|0100\rangle-|1011\rangle-|1101\rangle)$ \\\\
3. &$ \frac{a}{2}(|0000\rangle+|1111\rangle +|0011\rangle+1100\rangle) + 
\frac{b+1}{2}(|0101\rangle+|1010\rangle)$ \\ & $ 
+\frac{b-1}{2}(|0110\rangle+|1001\rangle)+\frac{1}{2}(|1101\rangle+|0010\rangle-
|0001\rangle-|1110\rangle)$ \\\\
6. 
&$\frac{a+b}{2}(|0000\rangle+|1111\rangle)+b(|0101\rangle+|1010\rangle)
+i(|1001\rangle-|0110\rangle)$ \\ & $ 
+\frac{a-b}{2}(|0011\rangle+|1100\rangle)+\frac{1}{2}(|0010\rangle+|0100\rangle+
|1011\rangle+|1101\rangle$ \\ & 
$-|0001\rangle-|0111\rangle-|1000\rangle-|1110\rangle)$ \\\\
9. 
&$a(|0000\rangle+|0101\rangle+|1010\rangle+|1111\rangle)$ \\ & $
-2i(|0100\rangle-|1001\rangle-|1110\rangle)$ \\\\
10.& $ \frac{a+i}{2}(|0000\rangle+|1111\rangle 
+|0011\rangle+1100\rangle)+\frac{a-i+1}{2}(|0101\rangle+|1010\rangle)$ \\ & $ 
+\frac{a-i-1}{2}(|0110\rangle+|1001\rangle)+\frac{i+1}{2}(|1101\rangle+
|0010\rangle)$ \\ & $
+\frac{i-1}{2}(|0001\rangle+|1110\rangle) 
-\frac{i}{2}(|0100\rangle+|0111\rangle+|1000\rangle+|1011\rangle)$ \\\\
12. 
&$(|0101\rangle-|0110\rangle+|1100\rangle+|1111\rangle)+(i+1)(|1001\rangle+|1010
\rangle)$ \\ & $-i(|0100\rangle+|0111\rangle+|1101\rangle-|1110\rangle)$ \\\\
14.& $ \frac{i+1}{2}(|0000\rangle+|1111\rangle-|0010\rangle-|1101\rangle)$ \\ & 
$ +\frac{i-1}{2}(|0001\rangle+|1110\rangle-|0011\rangle-1100\rangle)$ \\ & $ 
+\frac{1}{2}(|0100\rangle+|1001\rangle+|1010\rangle+|0111\rangle)$ \\ & 
$+\frac{1-2i}{2}(|1000\rangle+|0101\rangle+|0110\rangle+|1011\rangle)$ \\\\
16. 
&$\frac{1}{2}(|0\rangle+|1\rangle)\otimes(|000\rangle+|011\rangle+|100\rangle+|1
11\rangle$ \\ & $ +i(|001\rangle+|010\rangle-|101\rangle-|110\rangle))$ \\
\end{tabular}
\end{table}

We recall that $\SLloc^*$--orbits originating from different families are 
necessarily distinct, and that two states in the same family may be in the same 
$\SLloc^*$--orbit.

\newpage

\section*{Appendix B}

To simplify the notation, we shall use the same symbols to denote the 
polynomials in $\pA$ or $\pA^*$ and their restrictions to the Cartan subspace 
$\mfa$. As a byproduct of our construction of generators of the algebra $\pA^*$, 
we have obtained a nice set of generators of the algebra of polynomial 
invariants of the Weyl group of type $F_4$. Their degrees are, of course, 2, 6, 
8 and 12. The invariant of degree 12 has the factorization $\Pi=(L-M)(M-N)(N-L)$ 
and the one of degree 8 is a sum of three squares $\Sigma=L^2+M^2+N^2$ where 
\begin{tabular}{lll} $L$ & $=$ & $abcd$, \\ $M$ & $=$ & 
$\frac{1}{16}(4(ad-bc)^2-(a^2-b^2-c^2+d^2)^2)$ \\ & $=$ & 
$-\frac{1}{16}((a+b)^2-(c+d)^2)((a-b)^2-(c-d)^2)$, \\ $N$ & $=$ & 
$-\frac{1}{16}(4(ac+bd)^2-(a^2-b^2+c^2-d^2)^2)$ \\ & $=$ & 
$\frac{1}{16}((a+b)^2-(c-d)^2)((a-b)^2-(c+d)^2)$.\end{tabular} \\ The known sets 
of generators which we could find in the literature \cite{I,Me,SYS} do not share 
these special features.

\begin{table}[htbp]\caption{Generators $H$, $\Gamma$, $\Sigma$, $\Pi$ of 
invariants of the Weyl group of type $F_4$}\label{weylinv}
\begin{center}
\begin{tabular}{lll}
$2H$ & $=$ & $a^2+b^2+c^2+d^2$ \\
$2^5\Gamma$ & $=$ & $2(a^6+b^6+c^6+d^6)-(a^2+b^2+c^2+d^2)(a^4+b^4+c^4+d^4)$ \\ 
&& $+18(a^2b^2c^2+b^2c^2d^2+c^2d^2a^2+d^2a^2b^2)$ \\
$2^7\Sigma$ & $=$ & $2^8a^2b^2c^2d^2+(4(ad-bc)^2-(a^2-b^2-c^2+d^2)^2)$ \\ && 
$(4(ac+bd)^2-(a^2-b^2+c^2-d^2)^2)$ \\
$2^{12}\Pi$ & $=$ & $(16abcd-4(ad-bc)^2+(a^2-b^2-c^2+d^2)^2)$ \\ && 
$(4(ad-bc)^2-(a^2-b^2-c^2+d^2)^2$ \\ && 
$+4(ac+bd)^2-(a^2-b^2+c^2-d^2)^2)$ \\ && 
$(-4(ac+bd)^2+(a^2-b^2+c^2-d^2)^2-16abcd)$ \\
\end{tabular}
\end{center}
\end{table}

The generators $I_2$, $I_6$, $I_8$ and $I_{12}$ found in \cite{SYS} relate to 
our generators as follows:
\begin{enumerate}[]
\item $I_2=12H$,
\item $I_6=72H^3-96\Gamma$,
\item $I_8=264H^4-832\Gamma H +320\Sigma$,
\item $I_{12}=4104H^6-24096H^3 \Gamma+17440H^2 \Sigma+3904\Gamma^2-3840\Pi$.
\end{enumerate}


\clearpage

\begin{thebibliography}{99}

\providecommand{\bysame}{\leavevmode\hbox
to3em{\hrulefill}\thinspace}

\bibitem{BLT}
E. Briand, J.-G. Luque and J.-Y. Thibon, A complete set of covariants of the 
four qubit system, {\em J. Phys. A. : Math. Gen.} {\bf 38} (2003), 9915-9927.

\bibitem{BR}
H.J. Briegel and R. Raussendorf, Persistent entanglement in arrays of 
interacting particles, {\em Phys. Rev. Lett.} {\bf 86}, 910 (2000).

\bibitem{B}
J.-L. Brylinski, Algebraic measures of entanglement, {\em Chapter I in 
Mathematics of Quantum Computation}, Chapman and Hall/CRC, 2002, pp. 3-23.

\bibitem{BCS}
P. B\"{u}rgisser, M. Clausen and M. A. Shokrollahi, Algebraic Complexity Theory, 
Springer, 1997.

\bibitem{CH}
D. Choudhury and R. A. Horn, A complex orthogonal-symmetric analog of the polar 
decomposition, {\em SIAM J. Algebraic Discrete Methods} {\bf 8} (1987), 219-225.

\bibitem{CKW}
V. Coffman, J. Kundu and W. K. Wootters, Distributed entanglement, {\em Phys. 
Rev. A} {\bf 67}, 052306 (2000).

\bibitem{DW}
H. Derksen and J. Weyman, Generalized quivers associated to reductive groups, 
{\em Colloquium Mathematicum} {\bf 94} (2002), 151-173.

\bibitem{DLS}
D.\v{Z}. \DJ okovi\'{c}, N. Lemire and J. Sekiguchi, The closure ordering of 
adjoint nilpotent orbits in $\gso(p,q)$, {\em T\^{o}hoku Math. J.} {\bf 53} 
(2001), 395-442.

\bibitem{DZ}
D.\v{Z}. \DJ okovi\'{c} and K. Zhao, Tridiagonal normal forms for orthogonal 
similarity classes of symmetric matrices, {\em Linear Algebra and Its 
Applications} {\bf 384} (2004) 77--84.

\bibitem{DVC}
W. D\"{u}r, G. Vidal and J.I. Cirac, Three qubits can be entangled in two ways, 
{\em Phys. Rev. A} {\bf 62}, 062314 (2000).

\bibitem{G}
F. R. Gantmacher, Theory of Matrices, Vol. 2, Chelsea Publishing Company, 1959.

\bibitem{Ga}
[GAP] The GAP Group, GAP --- Groups, Algorithms, and Programming,
      Version 4.4.9; 2006
      (http://www.gap-system.org).

\bibitem{Si}
[GPS03] G.-M. Greuel, G. Pfister, and H. Sch\"onemann.
{\sc Singular} 2.05. A Computer Algebra System for Polynomial
Computations. Centre for Computer Algebra, University of
Kaiserslautern (2003). {\tt http://www.singular.uni-kl.de}.

\bibitem{H}
J.E. Humphreys, Reflection Groups and Coxeter Groups, Cambridge University 
Press, 1990.

\bibitem{I}
V.F. Ignatenko, Some questions in the geometric theory of invariants of groups 
generated by orthogonal and oblique reflections, {\em J. Soviet Math.} {\bf 33} 
(1986), 933-953.

\bibitem{K}
I. Kaplansky, Algebraic polar decomposition, {\em SIAM Journal on Matrix 
Analysis and Applications} {\bf 11} (1990), 213-217.

\bibitem{KP}
H. Kraft and C. Procesi, Closures of conjugacy classes of matrices are normal, 
{\em Inventiones Math.} {\bf 53} (1979), 227-247.

\bibitem{LLSS}
L. Lamata, J. Le\'{o}n, D. Salgado and  E. Solano, Inductive entanglement 
classification of four qubits under SLOCC, {\em arXiv: quant-ph/0610233 v.1}, 27 
Oct 2006.

\bibitem{LT}
J.-G. Luque and J.-Y. Thibon, Polynomial invariants of four qubits, {\em Phys. 
Rev. A} {\bf 67}, 042303 (2003).

\bibitem{LTT}
J.-G. Luque, J.-Y. Thibon and F. Toumazet, Unitary invariants of qubit systems, 
{\em arXiv: quant-ph/0604202 v1} 27 Apr 2006.

\bibitem{M}
Y. Makhlin, Nonlocal properties of two-qubit gates and mixed states, and 
optimization of quantum computation, {\em Quantum Information Processing} {\bf 
1} (2002), 243-252.

\bibitem{AM}
A. Mandilara, V.M. Akulin, A.V. Smilga and L.Viola, Quantum entanglement via 
nilpotent polynomials, {\em arXiv: quant-ph/0508234 v2} 15 Jan 2006.

\bibitem{Ma}
Maplesoft, Waterloo Maple Inc., Maple 9.01, 2003.

\bibitem{Me}
M. L. Mehta, Basic sets of invariant polynomials for finite reflection groups, 
{\em Comm. Algebra} {\bf 16} (1988), 1083-1098.

\bibitem{O}
T. Ohta, The closures of nilpotent orbits in the classical symmetric pairs and 
their singularities, {\em T\^{o}hoku Math. J.} {\bf 43} (1991), 161-211.

\bibitem{AP}
A.O. Pittenger, An Introduction to Quantum Computing Algorithms, Birkh\"{a}user, 
Boston, 2001.

\bibitem{JP}
J. Preskill, Lectures Notes for Physics 229: Quantum Information and 
Computation, California Institute of Technology, 1998.

\bibitem{SYS}
K. Saito, T. Yano and J. Sekiguchi, On a certain generator system of the ring of 
invariants of a finite reflection group, {\em Comm. Algebra} {\bf 8} (1980), 
373-408.

\bibitem{S}
L. Schl\"{a}fli, \"{U}ber die Resultante eines Systemes mehrerer algebraischer 
Gleichungen. Ein Beitrag zur Theorie der Elimination. {\em Denkschriften der 
Kaiserlichen Akademie Wissenschaften, mathematisch-naturwissenschaftliche 
Klasse} {\bf 4} (1852), Wien. Verzeichnis Grad, Nr. 24.
Reprinted in {\em Gesammelte mathematische Abhandlungen}, vol. 2,
Birkh\"auser, Basel, 1950-1956.

\bibitem{TY}
P. Tauvel and R.W.T. Yu, Lie Algebras and Algebraic Groups, Springer, 2005.

\bibitem{VDDV}
F. Verstraete, J. Dehaene, B. De Moor and H. Verschelde, Four qubits can be 
entangled in nine different ways,  {\em Phys. Rev. A} \mbox{\bf 65}, 052112 
(2002).

\bibitem{W}
N.R. Wallach, The Hilbert series of measures of entanglement for $4$ qubits, 
{\em Acta Applicandae Mathematicae} {\bf 86} (2005), 203-220.

\bibitem{WYKO}
C. Wu, Y. Yeo, L.C. Kwek and C.H. Oh, Quantum nonlocality of four-qubit 
entangled states, {\em arXiv: quant-ph/0611172 v1} 16 Nov 2006.
\end{thebibliography}
\end{document}